\documentstyle[12pt,epsf,epsfig]{article}
\newcount\timecount
\newcount\hours \newcount\minutes  \newcount\temp \newcount\pmhours

\hours = \time
\divide\hours by 60
\temp = \hours
\multiply\temp by 60
\minutes = \time
\advance\minutes by -\temp
\def\hour{\the\hours}
\def\minute{\ifnum\minutes<10 0\the\minutes
            \else\the\minutes\fi}
\def\clock{
\ifnum\hours=0 12:\minute\ AM
\else\ifnum\hours<12 \hour:\minute\ AM
      \else\ifnum\hours=12 12:\minute\ PM
            \else\ifnum\hours>12
                 \pmhours=\hours
                 \advance\pmhours by -12
                 \the\pmhours:\minute\ PM
                 \fi
            \fi
      \fi
\fi
}

\def\monthname{\relax\ifcase\month 0/\or January\or February\or
   March\or April\or May\or June\or July\or August\or September\or
   October\or November\or December\else\number\month/\fi}

\def\bold#1{\setbox0=\hbox{$#1$}%
     \kern-.025em\copy0\kern-\wd0
     \kern.05em\copy0\kern-\wd0
     \kern-.025em\raise.0433em\box0 }

\def\gappeq{\mathrel{\rlap {\raise.5ex\hbox{$>$}}
{\lower.5ex\hbox{$\sim$}}}}

\def\lappeq{\mathrel{\rlap{\raise.5ex\hbox{$<$}}
{\lower.5ex\hbox{$\sim$}}}}

\def\ga{\mathrel{\raise.3ex\hbox{$>$\kern-.75em\lower1ex\hbox{$\sim$}}}}
\def\la{\mathrel{\raise.3ex\hbox{$<$\kern-.75em\lower1ex\hbox{$\sim$}}}}
\def\gev{{\rm \, Ge\kern-0.125em V}}
\def\tev{{\rm \, Te\kern-0.125em V}}
\def\beq{\begin{equation}}
\def\eeq{\end{equation}}

\def\ohsq{\Omega_{\chi} h^2}

\def\m12{m_{1\!/2}}

\newcommand{\sLep}{\tilde{\ell}}
\newcommand{\sEl}{\tilde{e}}
\newcommand{\sMu}{\tilde{\mu}}
\newcommand{\sTau}{\tilde{\tau}}
\newcommand{\sNu}{\tilde{\nu}}
\newcommand{\chiz}{{\chi}^{0}}
\newcommand{\chipm}{{\chi}^{\pm}}

\begin{document}
\begin{titlepage}
\pagestyle{empty}
\baselineskip=21pt
\rightline{hep-ph/0106204}
\rightline{CERN--TH/2001-150}
\rightline{UMN--TH--2013/01, TPI--MINN--01/29}
\vskip 0.05in
\begin{center}
{\large{\bf Proposed Post-LEP Benchmarks for Supersymmetry
}}
\end{center}
\begin{center}
\vskip 0.05in
{{\bf Marco Battaglia}$^1$,
{\bf Albert De Roeck}$^1$,
{\bf John Ellis}$^1$,
{\bf Fabiola Gianotti}$^1$,\\
{\bf Konstantin T.~Matchev}$^1$,
{\bf Keith A.~Olive}$^{1,2}$,
{\bf Luc Pape}$^1$ and
{\bf Graham Wilson}$^3$
\vskip 0.05in
{\it
$^1${CERN, Geneva, Switzerland}\\
$^2${Theoretical Physics Institute, School of Physics and Astronomy,\\
University of Minnesota, Minneapolis, MN 55455, USA}\\
$^3${Department of of Physics and Astronomy, Schuster Laboratory,
University of Manchester, Manchester, UK}\\
}}
\vskip 0.05in
{\bf Abstract}
\end{center}
\baselineskip=18pt \noindent

We propose a new set of supersymmetric benchmark scenarios, taking into
account the constraints from LEP, $b \rightarrow s \gamma$, $g_\mu - 2$
and cosmology. We work in the context of the constrained MSSM (CMSSM) with
universal soft supersymetry-breaking masses and assume that $R$ parity is
conserved. We propose benchmark points that exemplify the different
generic possibilities, including focus-point models, points where
coannihilation effects on the relic density are important, and points with
rapid relic annihilation via direct-channel Higgs poles. We discuss the
principal decays and signatures of the different classes of benchmark
scenarios, and make initial estimates of the physics reaches of different
accelerators, including the Tevatron collider, the LHC, and $e^+ e^-$
colliders in the sub- and multi-TeV ranges. We stress the complementarity
of hadron and lepton colliders, with the latter favoured for
non-strongly-interacting particles and precision measurements. We mention
features that could usefully be included in future versions of
supersymmetric event generators. 

\vfill
\vskip 0.15in
\leftline{CERN--TH/2001-150}
\leftline{June 2001}
\end{titlepage}
\baselineskip=18pt

\section{Introduction}

The completion of the LEP experimental programme brings to an end an era
of precise electroweak measurements and the search for new particles 
with masses
$\lappeq 100$~GeV. With the start of Run II of the Fermilab
Tevatron collider, the advent of the LHC and perhaps a linear $e^+ e^-$
collider, the experimental exploration of the TeV energy scale is
beginning in earnest. 

The best-motivated scenario for new physics at the TeV energy scale is
generally agreed to be supersymmetry. Theoretically, it is compellingly
elegant, offers the possibility of unifying the fermionic matter particles
with the bosonic force particles, is the only framework thought to be
capable of connecting gravity with the other interactions, and appears
essential for the consistency of string theory. However, none of these
fundamental arguments offer clear advice as to the energy scale at which
supersymmetric particles might appear.

The first such argument was provided by the hierarchy problem: if
supersymmetric particles weigh less than of order 1~TeV, they may
stabilize the electroweak scale $m_Z \ll m_P \sim 10^{19}$~GeV. The
heavier the supersymmetric particles, the more the fine-tuning of the
model parameters required to fix $m_Z$ at its observed value. However, it
is difficult to attach quantitative significance to any specific
measure of the amount of this fine tuning.

LEP has provided no direct evidence for any physics beyond the Standard
Model, but it has provided several indirect hints that supersymmetry may
indeed lie around the corner. One such hint was provided by
LEP's
very accurate measurements of the gauge couplings, which are highly consistent
with a supersymmetric Grand Unified theory (GUT) if the supersymmetric
partners of the Standard Model particles weigh less than about
1~TeV~\cite{corner}, as
suggested by the hierarchy problem \cite{hierarchy}. Secondly, the precise
electroweak data from LEP and elsewhere suggest that the Higgs boson is
relatively light \cite{LEPEWWG}:
\begin{equation}
m_H \; =  \; 98^{+58}_{-38} \; {\rm GeV}
\label{Higgsmass}
\end{equation}
in good agreement with the prediction of the minimal supersymmetric
extension of the Standard Model (MSSM), if the squarks weigh $\lappeq
1$~TeV. Direct searches at LEP provided the lower limit $m_H >
113.5$~GeV (95\% CL)~\cite{LEPHiggs}. In
the final weeks of its run, LEP provided tentative evidence for
a Higgs boson weighing $115.0^{+1.3}_{-0.9}$~GeV \cite{LEPHiggs}, perfectly
consistent with the range (\ref{Higgsmass}) expected on the basis of the precise
LEP electroweak measurements, as well as with the MSSM. Indeed, the effective
potential of the Standard Model would be so sensitive to destabilization
by radiative corrections if this tentative LEP evidence were to be
confirmed, that some form of supersymmetry would probably be needed to
stabilize our familiar electroweak vacuum~\cite{ER}. 

A completely independent motivation for supersymmetry at the TeV scale is
provided by the cold dark matter advocated by astrophysicists and
cosmologists. If $R$ parity is conserved, 
as we assume here, the lightest supersymmetric
particle (LSP) is an ideal candidate to constitute the cold dark matter,
if it weighs $\lappeq 1$~TeV. 
We assume here that the LSP is the lightest neutralino $\chi$ \cite{EHNOS}.
The relic LSP density increases with the
masses of the supersymmetric particles, so the cosmological upper limit
$\ohsq \le 0.3$ may, in principle be used to set an upper limit on the
sparticle masses. However, in practice, one must be careful not to discard
`funnels' in the MSSM parameter space where heavier sparticles may be
permitted.

Finally, we should add that the recent precise measurement of the
anomalous magnetic moment of the muon, $g_\mu - 2$, which is in apparent
disagreement with the Standard Model at the 2.6-$\sigma$ level
\cite{Brown:2001mg}, has led to many speculations about new physics at the TeV
scale. Prominent among these have been various supersymmetric interpretations of
the possible discrepancy. These offer further encouragement that supersymmetry
might be discovered at the LHC or before. However, caution advises us to await
confirmation of the initial experimental value of $g_\mu - 2$ and to seek
consensus on the calculation of hadronic contributions to $g_\mu - 2$
before jumping to any conclusions.

Nevertheless, the front-running nature of the supersymmetric candidacy for
new physics beyond the Standard Model has motivated many studies of its
experimental signatures at future colliders. In order to focus these
discussions, and to provide standards of comparison for different
analyses, experiments and accelerators, specific benchmark choices of
supersymmetric parameters have often been proposed. For example, several
years ago, several such benchmark scenarios were used to evaluate the
capabilities of the LHC for detecting
supersymmetry~\cite{Paige,ATLASTDR,CMS98006}. More recently, analogous
benchmarks have been used in linear collider
studies~\cite{TESLATDR}. 

Unfortunately, time has overtaken some of these benchmark scenarios, whose
parameters have by now been excluded by direct experimental searches for
superymmetry and the Higgs boson at LEP, or because they
predict unacceptable values of $g_\mu - 2$ or $b \rightarrow s \gamma$
decay, or because they predict unacceptable values of the relic LSP
density $\ohsq$.

The purpose of this paper is to propose a new set of benchmark
supersymmetric model parameters that are consistent with the experimental
constraints, as well as cosmology. They may
therefore provide helpful aids for understanding better the
complementarity of different accelerators in the TeV energy range. We
restrict our attention to a constrained version of the MSSM (CMSSM) which
incorporates a minimal supergravity (mSUGRA)-inspired model of soft
supersymmetry breaking. In the CMSSM, universal gaugino masses $m_{1/2}$,
scalar masses $m_0$ (including those of the Higgs multiplets) and
trilinear supersymmetry breaking parameters $A_0$ are input at the
supersymmetric grand unification scale. In this framework, the Higgs
mixing parameter $\mu$ can be derived (up to a sign) from the other MSSM
parameters by imposing the electroweak vacuum conditions for any given
value of $\tan \beta$.  Thus, given the set of input parameters determined
by $\{ m_{1/2}, m_0, A_0,\tan\beta,sgn(\mu) \}$, the entire spectrum of
sparticles can be derived.  Here we will further restrict our attention to
$A_0 = 0$, for simplicity. We do not consider benchmarks for models with
gauge-
\cite{Dine}, gaugino-
\cite{Kaplan} or anomaly-mediated
\cite{Randall} supersymmetry breaking, or for models
with broken $R$ parity. Studies of these and other models would be interesting
complements to this work, and we comment on them in the last Section of
this paper. 

\begin{figure}
\epsfig{file=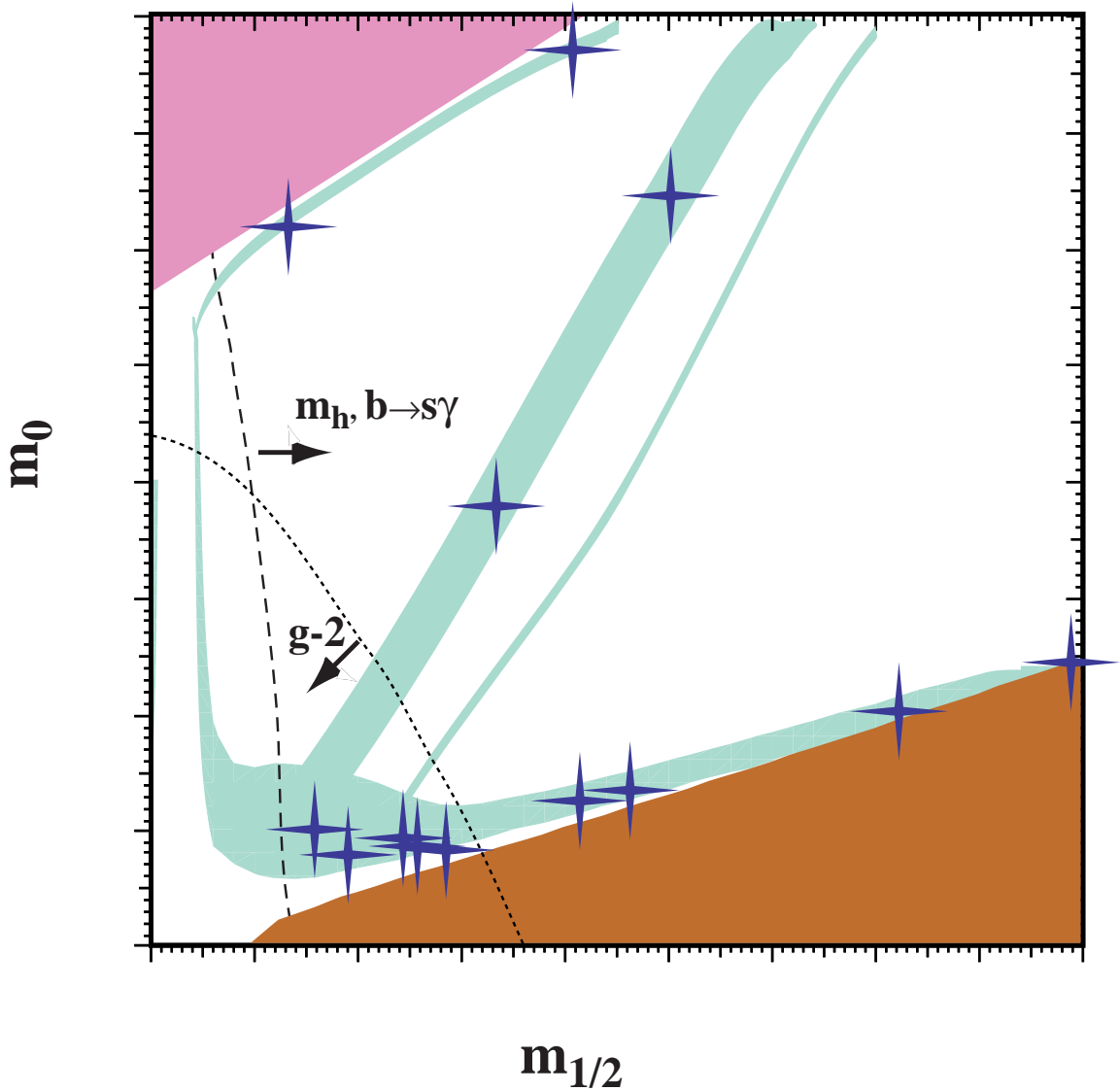,height=6in}
\caption{\label{fig:rough}
{\it 
Qualitative overview of the locations of our proposed benchmark points in
a generic $(m_{1/2}, m_0)$ plane.  The light (turquoise) shaded area is
the cosmologically preferred region with \protect\mbox{$0.1\leq\ohsq\leq
0.3$}, whose exact shape depends on the value of $\tan\beta$, and to some
extent on the Standard Model inputs $m_t$, $m_b$ and $\alpha_s$.  In the
dark (brick red) shaded region at bottom right, the LSP is the charged
${\tilde \tau}_1$, so this region is excluded. Electroweak symmetry
breaking is not possible in the dark (pink) shaded region at top left. The
LEP experimental constraints, in particular that on $m_h$, and
measurements of $b \rightarrow s \gamma$ exert pressure from the left
side. The BNL E821 measurement of $g_\mu - 2$ favours relatively low
values of $m_0$ and $m_{1/2}$ for $\mu > 0$. The CMSSM benchmark points we
propose are indicated roughly by the (blue) crosses. We propose points in
the `bulk' region at bottom left, along the coannihilation `tail'
extending to larger $m_{1/2}$, in the `focus-point' region at large $m_0$,
and in the rapid-annilation `funnel' that may appear at intermediate $m_0
/ m_{1/2}$ for large $\tan \beta$. 
}}
\end{figure}

Fig.~\ref{fig:rough} illustrates qualitatively the CMSSM benchmark points
we propose, superimposed on the regions of the $(m_{1/2}, m_0)$ plane
favoured by LEP limits, particularly on $m_h$, $b \rightarrow s \gamma$
and cosmology. Electroweak symmetry breaking is not possible in the
dark-shaded triangular region in the top left corner, and the lightest
supersymmetric particle would be charged in the bottom right dark-shaded
triangular region. The experimental constraints on $m_h$ and $b
\rightarrow s \gamma$ exert pressures from the left, as indicated, which
depend on the value of $\tan \beta$ and the sign of $\mu$. The indication
of a deviation from the Standard Model in $g_\mu - 2$ disfavours $\mu < 0$
and large values of $m_0$ and $m_{1/2}$ for $\mu > 0$. The region where
$\ohsq$ falls within the preferred range is indicated in light shading,
its exact shape being dependent on the value of $\tan\beta$, and to some
extent on the Standard Model inputs $m_t$, $m_b$ and $\alpha_s$.  As
discussed later in more detail, in addition to the `bulk' region at low
$m_0$ and $m_{1/2}$, there is a coannihilation `tail' extending to large
$m_{1/2}$~\cite{EFOSi,glp}, a `focus-point' region at large $m_0$ near the
boundary of the region with proper electroweak symmetry
breaking~\cite{Feng:2000gh}, and narrow rapid-annilation `funnels' at
intermediate $m_0 / m_{1/2}$ for large $\tan
\beta$~\cite{dn,BB,Baer,lns,EFGOSi}. The interplays of these features for
different values of $\tan \beta, sgn(\mu)$, together with the
corresponding proposed benchmark points, are shown in
Figs.~\ref{fig:locations1} and \ref{fig:locations2}. 

It is possible to learn much from general theoretical scans of the CMSSM
parameter space, but the LHC experience also showed the complementary
advantages of devoting some experimental attention to specific benchmark
points~\cite{Paige}, where the nitty-gritty problems of disentangling
possible overlapping experimental signals and extracting measurements of
CMSSM parameters can be confronted. We do not propose here a `fair'
statistical sampling of the allowed CMSSM parameter space. Rather, we
propose benchmark points that span the essential range of theoretical
possibilities, given our present knowledge. Some of the points we propose
might soon become obsolete, for example because of Tevatron data or
reductions in the error in $g_\mu - 2$. As seen in Fig.~\ref{fig:rough},
many of the points we propose are spread over the allowed part of the
`bulk' region, at different values of $\tan \beta$.  However, we also
propose some points strung along the coannihilation `tail', including one
at the extreme tip, and two points each in the `focus-point' region and
the rapid-annihilation `funnels'. Some of these points might appear
disfavoured by fine-tuning arguments~\cite{EENZ,EO}, but cannot be
excluded. Taken together, the points we propose serve to highlight the
different possible scenarios with which future colliders may be
confronted. The input parameters for the benchmark points we propose,
together with the resulting spectra as calculated using the code {\tt
SSARD}~\cite{SSARD}, are shown in Table~\ref{tab:msugra.spectr_Olive}.

\begin{table}[p!]
\centering
\renewcommand{\arraystretch}{0.95}
{\bf Supersymmetric spectra}\\
{~}\\
\begin{tabular}{|c||r|r|r|r|r|r|r|r|r|r|r|r|r|}
\hline
Model          & A   &  B  &  C  &  D  &  E  &  F  &  G  &  H  &  I  &  J  &  K  &  L  &  M   \\ 
\hline
$m_{1/2}$      & 600 & 250 & 400 & 525 &  300& 1000& 375 & 1500& 350 & 750 & 1150& 450 & 1900 \\
$m_0$          & 140 & 100 &  90 & 125 & 1500& 3450& 120 & 419 & 180 & 300 & 1000& 350 & 1500 \\
$\tan{\beta}$  & 5   & 10  & 10  & 10  & 10  & 10  & 20  & 20  & 35  & 35  & 35  & 50  & 50   \\
sign($\mu$)    & $+$ & $+$ & $+$ & $-$ & $+$ & $+$ & $+$ & $+$ & $+$ & $+$ & $-$ & $+$ & $+$  \\ 
$\alpha_s(m_Z)$& 120 & 123 & 121 & 121 & 123 & 120 & 122 & 117 & 122 & 119 & 117 & 121 & 116 \\
$m_t$          & 175 & 175 & 175 & 175 & 171 & 171 & 175 & 175 & 175 & 175 & 175 & 175 & 175  \\ \hline
Masses         &     &     &     &     &     &     &     &     &     &     &     &     &      \\ \hline
$|\mu(m_Z) |$  & 739 & 332 & 501 & 633 & 239 & 522 & 468 &1517 & 437 & 837 &1185 & 537 & 1793 \\ \hline
h$^0$          & 114 & 112 & 115 & 115 & 112 & 115 & 116 & 121 & 116 & 120 & 118 & 118 &  123 \\
H$^0$          & 884 & 382 & 577 & 737 &1509 &3495 & 520 &1794 & 449 & 876 &1071 & 491 & 1732 \\
A$^0$          & 883 & 381 & 576 & 736 &1509 &3495 & 520 &1794 & 449 & 876 &1071 & 491 & 1732 \\
H$^{\pm}$      & 887 & 389 & 582 & 741 &1511 &3496 & 526 &1796 & 457 & 880 &1075 & 499 & 1734 \\ \hline
$\chi^0_1$     & 252 &  98 & 164 & 221 & 119 & 434 & 153 & 664 & 143 & 321 & 506 & 188 &  855 \\
$\chi^0_2$     & 482 & 182 & 310 & 425 & 199 & 546 & 291 &1274 & 271 & 617 & 976 & 360 & 1648 \\
$\chi^0_3$     & 759 & 345 & 517 & 654 & 255 & 548 & 486 &1585 & 462 & 890 &1270 & 585 & 2032 \\
$\chi^0_4$     & 774 & 364 & 533 & 661 & 318 & 887 & 501 &1595 & 476 & 900 &1278 & 597 & 2036 \\
$\chi^{\pm}_1$ & 482 & 181 & 310 & 425 & 194 & 537 & 291 &1274 & 271 & 617 & 976 & 360 & 1648 \\
$\chi^{\pm}_2$ & 774 & 365 & 533 & 663 & 318 & 888 & 502 &1596 & 478 & 901 &1279 & 598 & 2036 \\ \hline
$\tilde{g}$    &1299 & 582 & 893 &1148 & 697 &2108 & 843 &3026 & 792 &1593 &2363 & 994 & 3768 \\ \hline
$e_L$, $\mu_L$ & 431 & 204 & 290 & 379 &1514 &3512 & 286 &1077 & 302 & 587 &1257 & 466 & 1949 \\
$e_R$, $\mu_R$ & 271 & 145 & 182 & 239 &1505 &3471 & 192 & 705 & 228 & 415 &1091 & 392 & 1661 \\
$\nu_e$, $\nu_{\mu}$
               & 424 & 188 & 279 & 371 &1512 &3511 & 275 &1074 & 292 & 582 &1255 & 459 & 1947 \\
$\tau_1$       & 269 & 137 & 175 & 233 &1492 &3443 & 166 & 664 & 159 & 334 & 951 & 242 & 1198 \\
$\tau_2$       & 431 & 208 & 292 & 380 &1508 &3498 & 292 &1067 & 313 & 579 &1206 & 447 & 1778 \\
$\nu_{\tau}$   & 424 & 187 & 279 & 370 &1506 &3497 & 271 &1062 & 280 & 561 &1199 & 417 & 1772 \\ \hline
$u_L$, $c_L$   &1199 & 547 & 828 &1061 &1615 &3906 & 787 &2771 & 752 &1486 &2360 & 978 & 3703 \\
$u_R$, $c_R$   &1148 & 528 & 797 &1019 &1606 &3864 & 757 &2637 & 724 &1422 &2267 & 943 & 3544 \\
$d_L$, $s_L$   &1202 & 553 & 832 &1064 &1617 &3906 & 791 &2772 & 756 &1488 &2361 & 981 & 3704 \\
$d_R$, $s_R$   &1141 & 527 & 793 &1014 &1606 &3858 & 754 &2617 & 721 &1413 &2254 & 939 & 3521 \\
$t_1$          & 893 & 392 & 612 & 804 &1029 &2574 & 582 &2117 & 550 &1122 &1739 & 714 & 2742 \\
$t_2$          &1141 & 571 & 813 &1010 &1363 &3326 & 771 &2545 & 728 &1363 &2017 & 894 & 3196 \\ 
$b_1$          &1098 & 501 & 759 & 973 &1354 &3319 & 711 &2522 & 656 &1316 &1960 & 821 & 3156 \\
$b_2$          &1141 & 528 & 792 &1009 &1594 &3832 & 750 &2580 & 708 &1368 &2026 & 887 & 3216 \\ \hline
\hline
\end{tabular}
\caption[]{\it Proposed CMSSM benchmark points and
mass spectra (in GeV), as calculated using {\tt SSARD}~\cite{SSARD} 
and {\tt FeynHiggs}~\cite{Heinemeyer:2000yj}. 
The renormalization-group equations are run down to the
electroweak scale $m_Z$, where the one-loop corrected effective potential
is computed and the CMSSM spectroscopy calculated, including the
one loop corrections to the chargino and neutralino masses.
The pseudoscalar Higgs mass $m_A$ is computed as in~\cite{Carena:2000yi}. 
Exact gauge coupling unification is enforced and the prediction for
$\alpha_s(m_Z)$ is shown (in units of 0.001). 
It is also assumed that $A_0 = 0$ and $m_b(m_b)^{\overline {MS}} = 4.25$~GeV.
For most of the points, $m_t = 175$~GeV is used, but for points E and F
the lower value $m_t = 171$~GeV is used, for better consistency
with~\cite{Feng:2000gh}.} 
\label{tab:msugra.spectr_Olive}
\end{table}

The layout of this paper is as follows. In Section 2, we discuss the
various experimental and other constraints on supersymmetric scenarios,
and discuss how we implement them. Then, in Section 3, we introduce the
set of benchmark scenarios we propose, motivating our choices in the
multidimensional parameter space of the MSSM. For convenience, we
introduce in Section 4 versions of these benchmarks calculated with
suitable {\tt ISASUGRA}~\cite{isasugra} inputs, and we then use {\tt
ISASUGRA} to discuss the decay signatures of heavier sparticles, which are
quite distinctive in some of these benchmark scenarios. Then, in Section 5
we take first looks at the physics reaches of various TeV-scale colliders,
including the Tevatron, the LHC, a 500-GeV to 1-TeV linear $e^+ e^-$
collider such as TESLA, the NLC or the JLC, and a 3- to 5-TeV linear $e^+
e^-$ collider such as CLIC~\footnote{We comment in passing on the
capabilities of $\mu^+ \mu^-$ colliders.}.  Finally, in Section 6, we
review our results on the CMSSM benchmark scenarios we propose, discuss
some of the future work that might be done to investigate further these
benchmark supersymmetric scenarios and use them as a guide to
understanding the physics opportunities offered by future colliders, and
mention other possible scenarios that could also be studied. 

\section{Experimental and Cosmological Constraints}

We implement the experimental and cosmological constraints using a code
{\tt SSARD} that incorporates the two-loop running of the input soft
supersymmetry-breaking parameters from the input scale $M_{\rm GUT}$
(defined as the scale where $g_1$ and $g_2$ meet) down to the
electroweak scale, identified with $m_Z$. Exact gauge coupling unification
is enforced, and the strong couplings constant $\alpha_s(m_Z)$
is a prediction. The $\mu$ parameter is extracted by minimizing the
one-loop corrected effective potential \cite{CMSSM,Barger:1994gh}
at the scale $m_Z$, while the pseudoscalar Higgs mass
$m_A$ is computed using the results of \cite{Carena:2000yi}.
The radiative corrections to the light Higgs boson mass $m_h$ are
computed with the {\tt FeynHiggs} code \cite{Heinemeyer:2000yj}.
The full one-loop corrections to the physical chargino and neutralino masses
are included \cite{Pierce:1994gj,Pierce:1994ew,Pierce:1997zz}.
The code also calculates $b \to s \gamma$~\cite{gerri}, $g_\mu - 2$
and the cosmological relic density using consistent conventions. We note, in
particular, that the inclusion of one-loop corrections to chargino and
neutralino masses is important for implementing accurately the LEP limits and
the boundary of the region favoured by cosmology. We discuss later the problems
encountered in matching the sparticle spectra obtained using this and
other codes that implement these constraints using different
approaches and/or approximations.

\subsection{Sparticle Searches}

The most important direct experimental constraints on the MSSM parameter
space are provided by LEP searches for sparticles~\cite{SUSYWG} and Higgs
bosons~\cite{LEPHiggs}, the latter constraining the sparticle spectrum
indirectly via radiative corrections, particularly those associated with
third-generation supermultiplets. We use here the preliminary combined
results that are based on data-taking at centre-of-mass energies up to
about 208 GeV.

Upper limits at 95\% CL on the the cross section for chargino-pair
production were set \cite{LEPSUSYWG_0103} for all kinematically accessible
chargino masses as a function of the neutralino mass, assuming that the
branching ratio for ${\chi^\pm} \rightarrow {W^\pm} {\chi^0}$ was 100\%. 
For neutralino masses approximately half the chargino mass, the upper
limit obtained using 35 pb$^{-1}$ of integrated luminosity at $\sqrt{s} >
207.5 GeV$ is around 0.5 pb. These cross-section limits can be interpreted
within the MSSM for some specific parameter values; for $\tan{\beta}=2$,
$\mu=-200$ GeV and sneutrino masses exceeding 300 GeV, the lower limit on
the chargino mass is 103.5 GeV \cite{LEPSUSYWG_0103}~\footnote{There are
also model-dependent limits on the supersymmetric parameter space derived
from searches for associated $\chi \chi'$ production at LEP.}. 

Similarly, the combined LEP data at $\sqrt{s}$ from 183 to 208 GeV were
used to search for sleptons~\cite{LEPSUSYWG_0101}.  Events containing two
charged leptons and missing energy were analysed and upper limits set on
the cross section times branching-ratio squared for slepton-pair
production followed by ${\tilde \ell} \rightarrow \ell + \chi$ decay, as
functions of the slepton and neutralino masses.  The limits vary
substantially with the masses, but typically the limits are 40~fb for the
selectron and smuon search and 100~fb for the stau search.  Within the
context of the MSSM, these experimental limits lead to the exclusion of
major portions of the right-handed slepton, neutralino mass plane at 95\%
CL.  The mass limits were evaluated for $\tan{\beta}=1.5$ and $\mu=-200$
GeV. For a neutralino mass of 40 GeV, the lower limits on the right handed
slepton masses are 99.4~GeV, 96.4~GeV and 87.1~GeV for the selectron,
smuon and stau respectively~\footnote{We note that,
for both the chargino and the slepton searches, the
sensitivity is much reduced for small values of the mass difference
between the parent sparticle and the neutralino. However, this caveat is
unimportant for the CMSSM as studied here.}.

There are also important constraints on the squark and gluino masses from
Run I of the Tevatron~\cite{Spiro}, extending up to 300~GeV if $m_{\tilde
q} \sim m_{\tilde g}$, and additional constraints on stop and sbottom
squarks from LEP. Since they do not play a r\^ole in our analysis, we do
not discuss them in detail. 

\begin{figure}
\vspace*{-0.75in}
\begin{minipage}{7.5in}
\epsfig{file=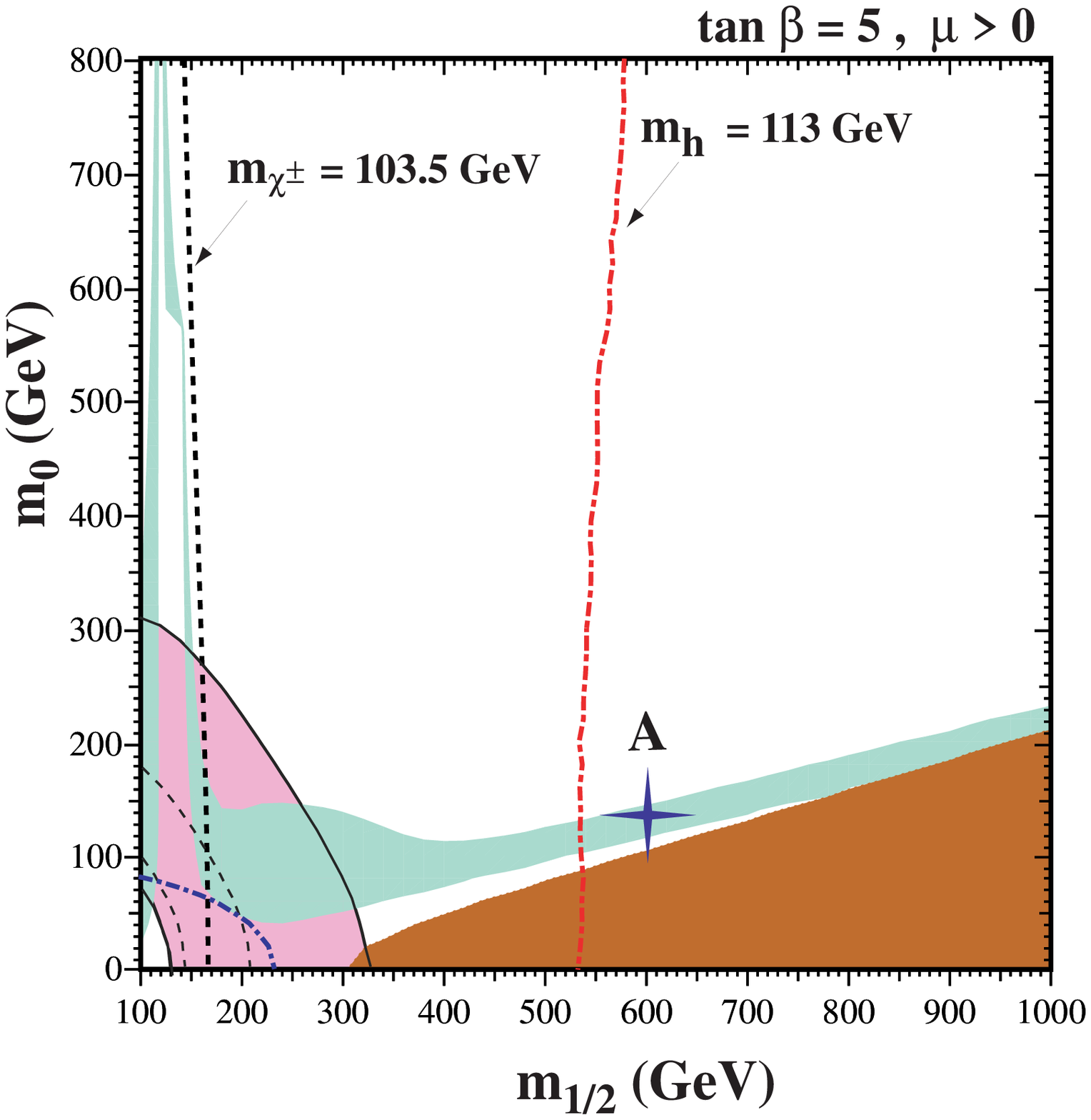,height=3.25in}
\epsfig{file=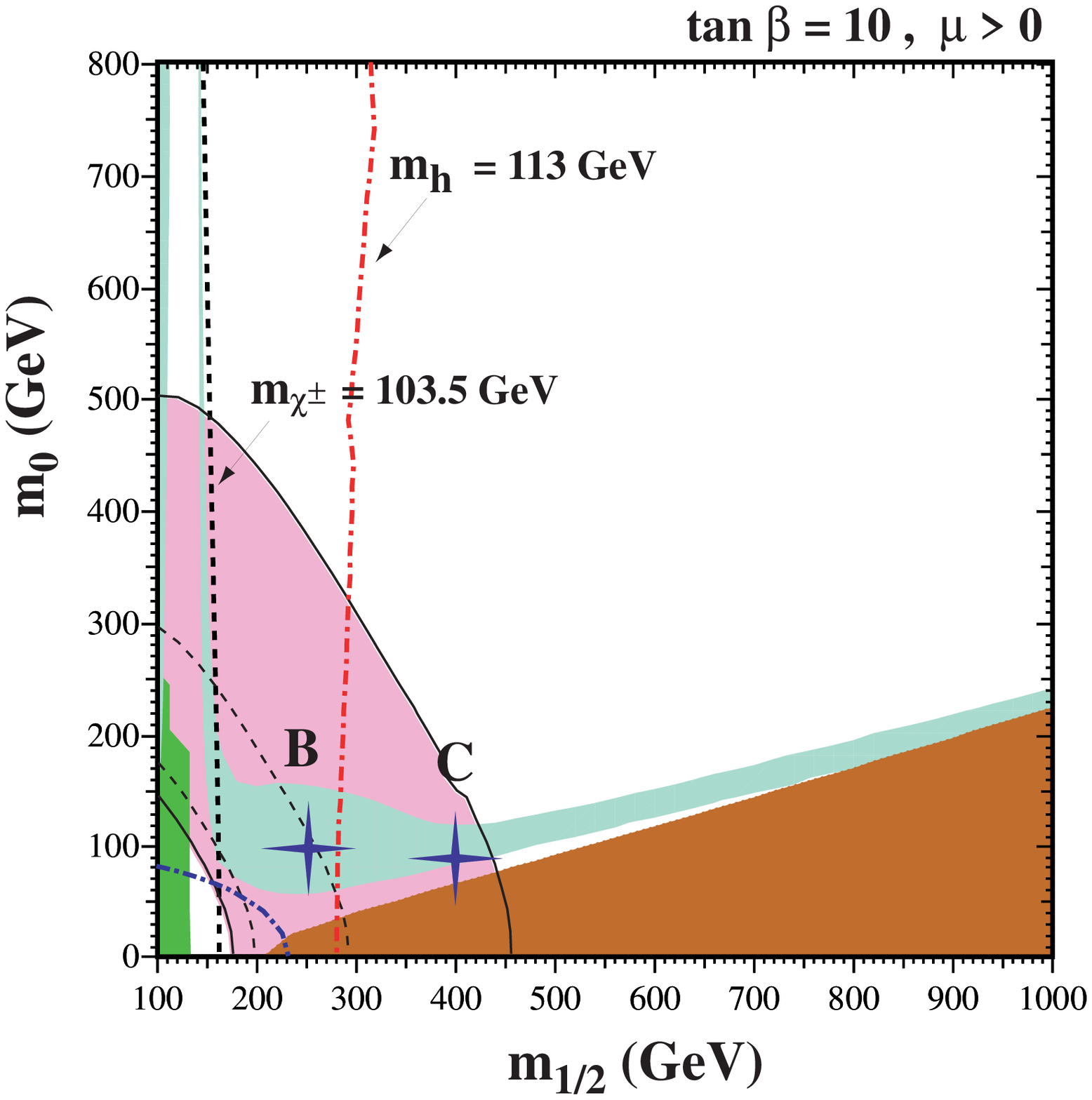,height=3.25in} \hfill
\end{minipage}
\begin{minipage}{7.5in}
\epsfig{file=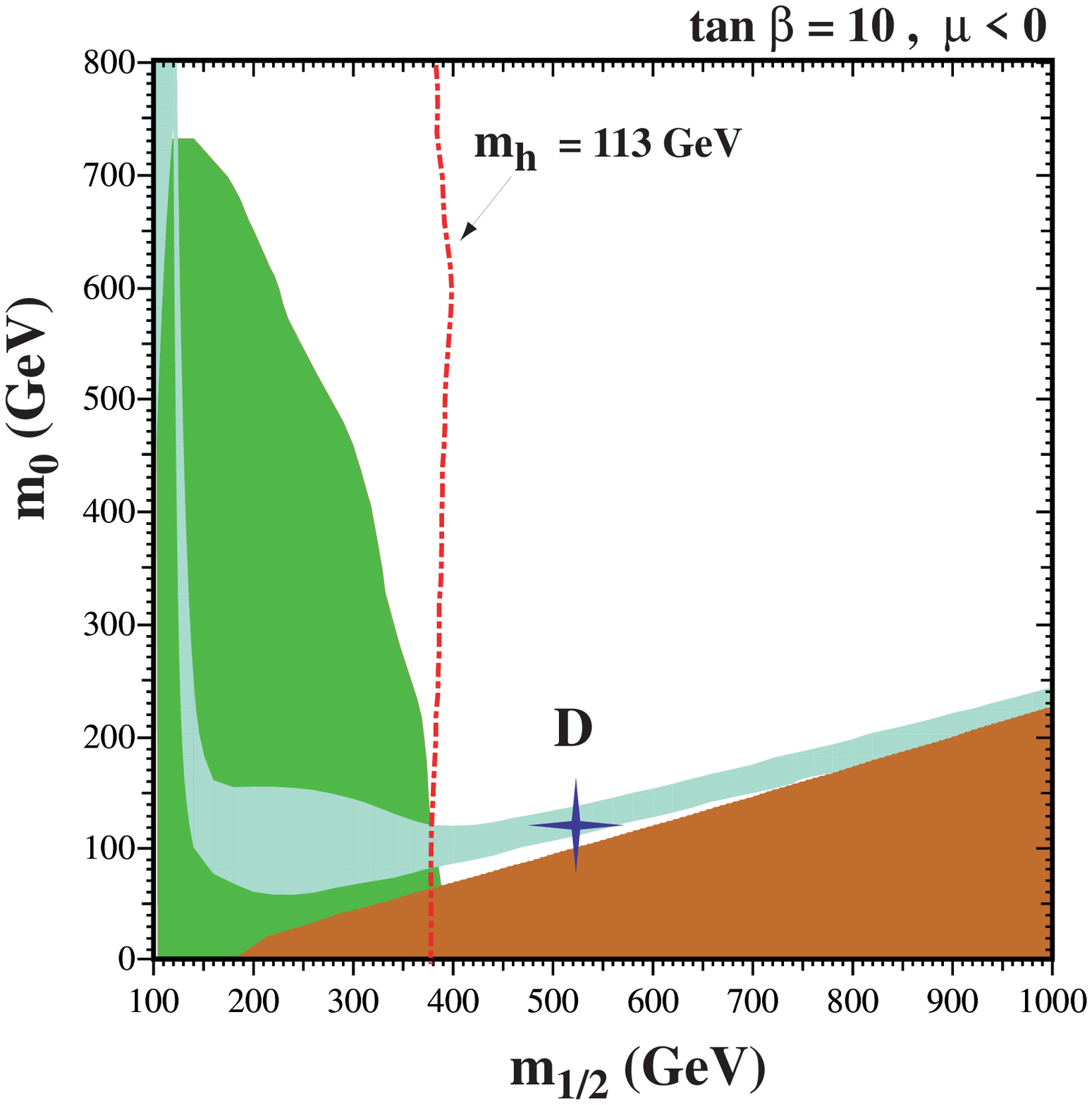,height=3.25in}
\epsfig{file=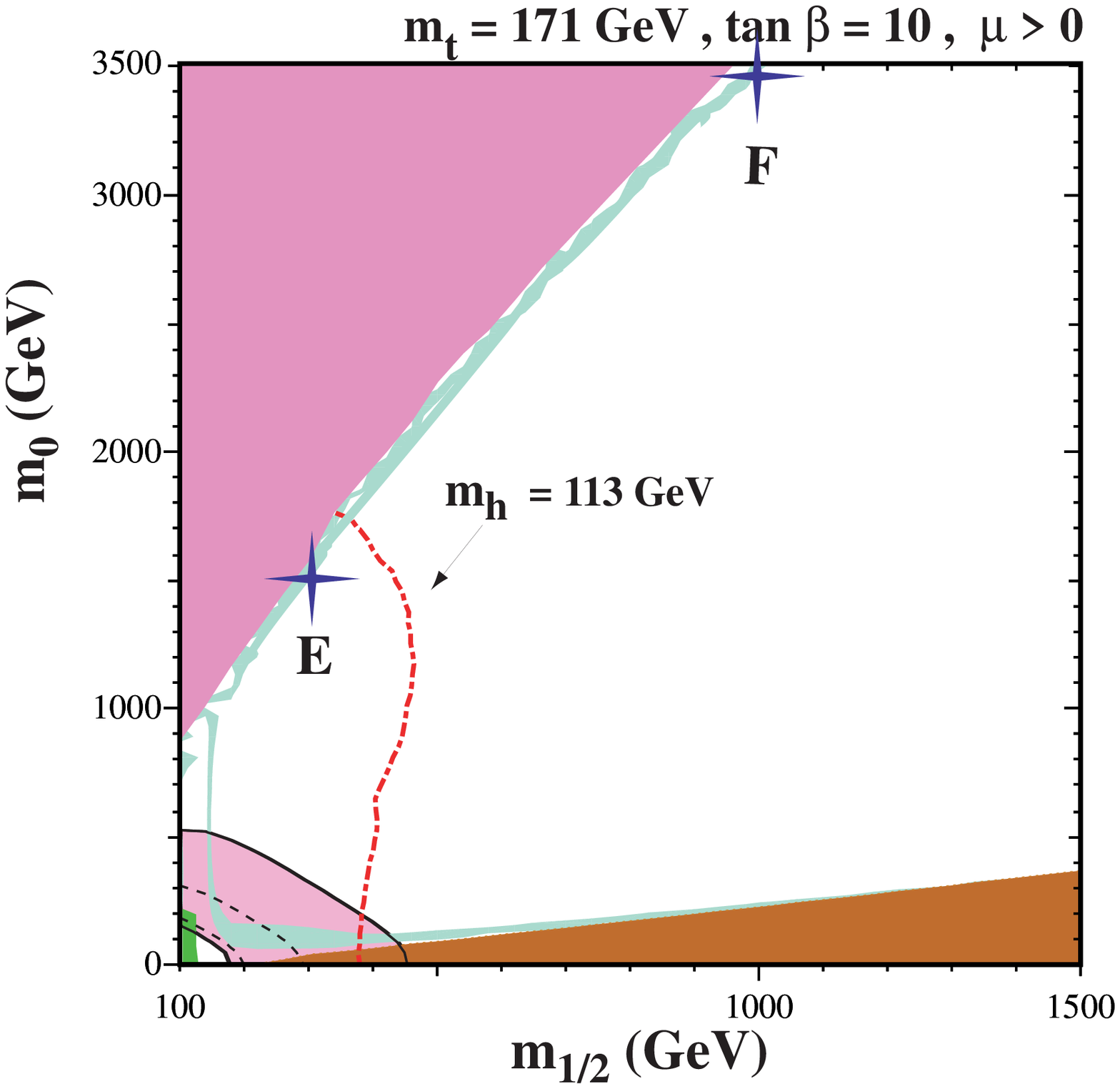,height=3.25in} \hfill
\end{minipage}
\caption{\label{fig:locations1}
{\it The $(m_{1/2}, m_0)$ planes for $\tan \beta =$ (a) 5 ($\mu > 0$),
(b) 10 ($\mu > 0$), (c) 10 ($\mu < 0$), all for $m_t = 175$~GeV,
and (d) 10
($\mu > 0$) with $m_t = 171$~GeV. In each case we have assumed $A_0 = 0$ and  
$m_b(m_b)^{\overline {MS}}_{SM} = 4.25$~GeV, and used the {\tt SSARD}
code. The near-vertical (red) dot-dashed lines are the contours
$m_h = 113$~GeV, as evaluated using the {\tt FeynHiggs} code.
The medium (dark green) shaded regions are excluded by $b
\to s \gamma$.
The light (turquoise) shaded areas are the cosmologically
preferred
regions with \protect\mbox{$0.1\leq\ohsq\leq 0.3$}. In the
dark (brick red) shaded regions, the LSP is the charged ${\tilde \tau}_1$,
so this region is excluded. The regions allowed by the E821 measurement of
$a_\mu$ at the 2-$\sigma$ level are shaded (pink) and bounded by solid
black lines, with dashed lines indicating the 1-$\sigma$ ranges. 
Electroweak symmetry
breaking is not possible in the dark (pink) shaded region at the top left 
of
panel (d). The (blue) crosses denote the proposed benchmark points A to F.}}
\end{figure}

Analyses indicate that the experimental search results can be interpreted
as chargino and slepton mass bounds close to the kinematic limits for most
of the CMSSM parameter range~\cite{EFGOS}. Therefore, we show in
panels (a,b) of Fig.~\ref{fig:locations1} the contours (dot-dashed) in the
$(m_{1/2}, m_0)$ plane corresponding to $m_{\chi^\pm} = 103.5$~GeV and
$m_{\tilde e} = 99$~GeV.  These contours are omitted from the remaining
figures, for clarity. 


\begin{figure}
\vspace*{-0.75in}
\begin{minipage}{7.5in}
\epsfig{file=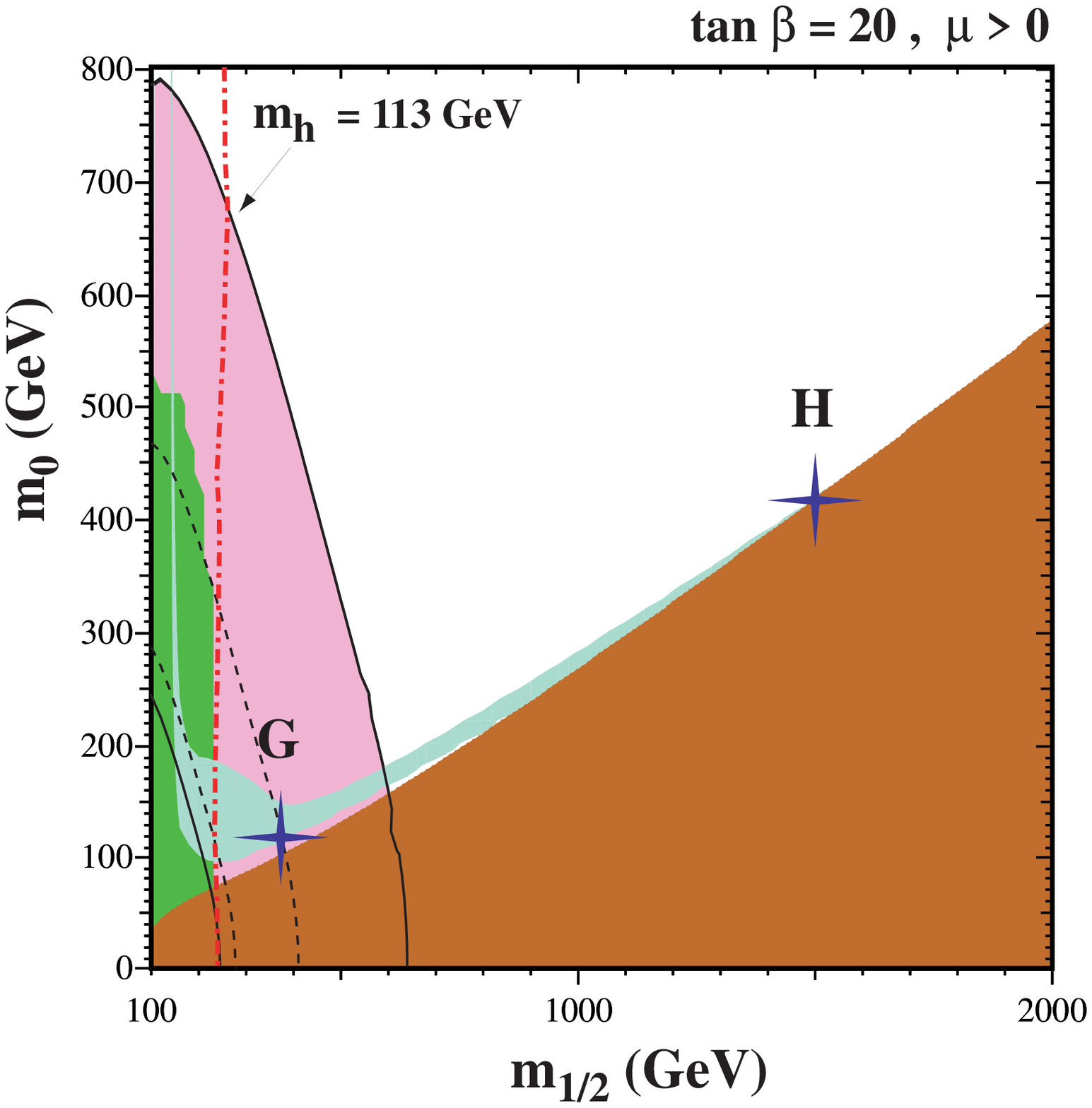,height=3.25in}
\epsfig{file=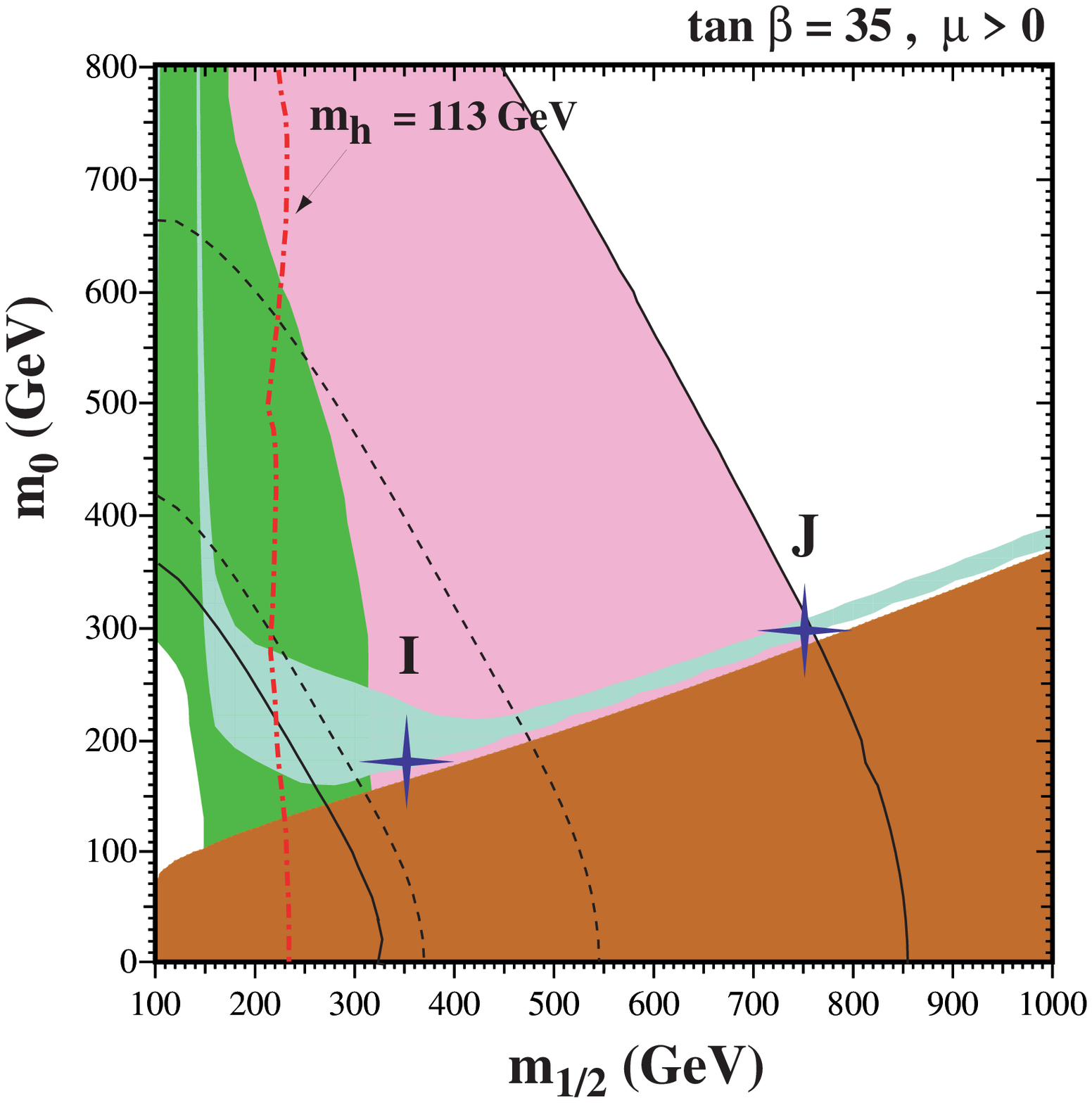,height=3.25in} \hfill
\end{minipage}
\begin{minipage}{7.5in}
\epsfig{file=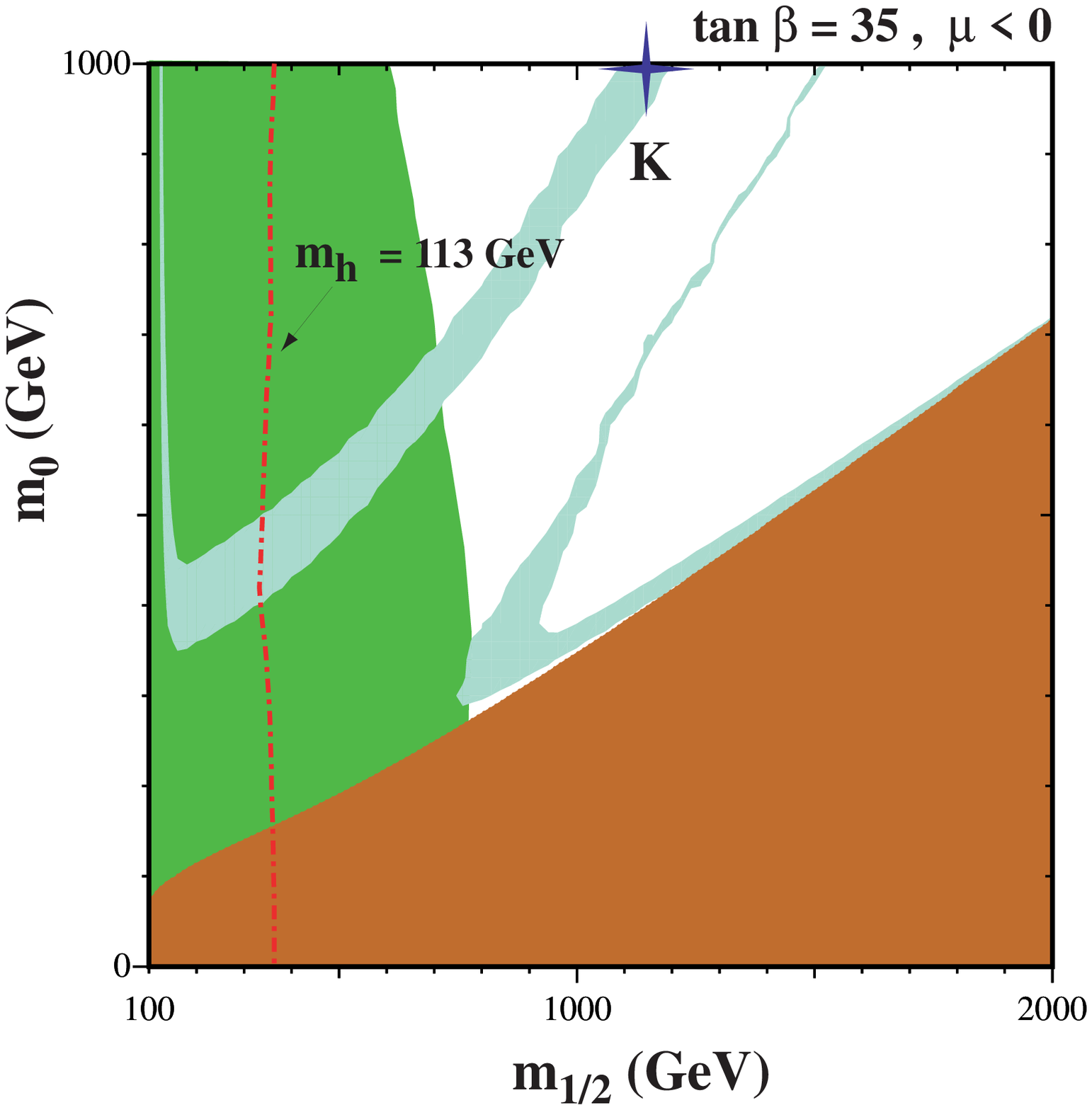,height=3.25in}
\epsfig{file=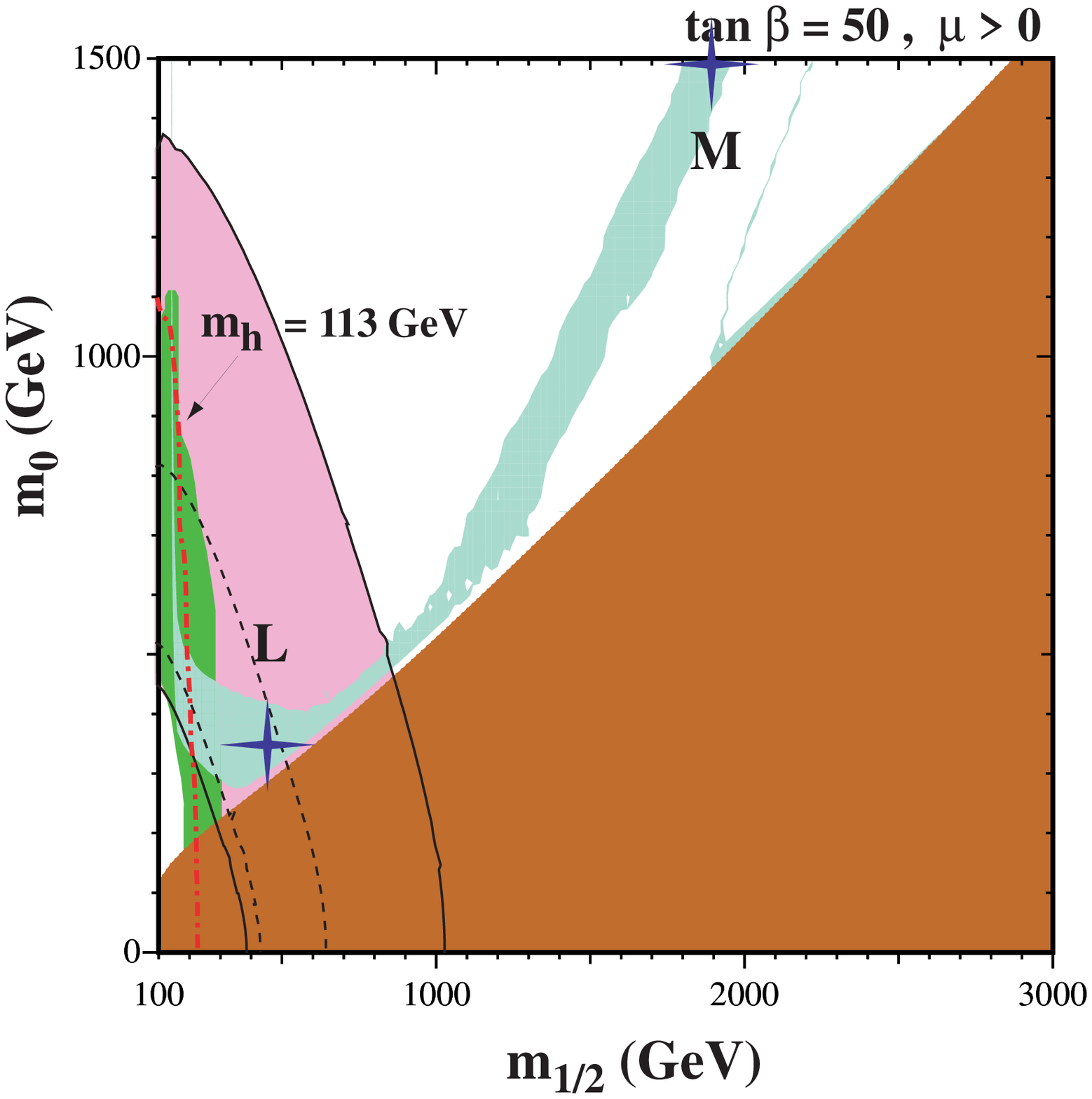,height=3.25in} \hfill
\end{minipage}
\caption{\label{fig:locations2}
{\it The $(m_{1/2}, m_0)$ planes for $\tan \beta =$ (a) 20 ($\mu > 0$),
(b) 35 ($\mu > 0$), (c) 35 ($\mu < 0$), and (d) 50
($\mu > 0$), found using {\tt SSARD} and assuming $A_0 = 0, m_t =
175$~GeV and   
$m_b(m_b)^{\overline {MS}}_{SM} = 4.25$~GeV. The notations are the same as
in Fig.~\ref{fig:locations1}.
The (blue) crosses denote the proposed benchmark points G to M. 
At larger $\tan\beta$, the size as well as the
exact shape of the cosmologically preferred region obtained
is subject to considerable uncertainty, and different 
programs yield different answers for the same fixed
values of the input parameters. The differences arise 
due to different calculational algorithms, and to
neglecting different sets of higher-order terms.
We elaborate more on these issues in Section~\ref{uncertainties}.
}}
\end{figure}

As mentioned in the Introduction, the lower limit on the mass of a
Standard Model Higgs boson imposed by the combined LEP experiments is
$113.5$~GeV~\cite{LEPHiggs}. This lower limit applies also to the MSSM for
small $\tan \beta$, even if squark mixing is maximal. In the CMSSM,
maximal mixing is not attained, and the $e^+ e^- \to Z^0 + h$ production
rate is very similar to that in the Standard Model, for all
values of $\tan \beta$. Therefore, the LEP hint for a Higgs boson
weighing $115.0^{+1.3}_{-0.9}$~GeV~\cite{LEPHiggs}, which is compatible 
with the background-only hypothesis at the 0.4\% CL, can also be interpreted
in the CMSSM. 

To calculate theoretically the mass of the lightest MSSM Higgs boson, we
use the {\tt FeynHiggs} code~\cite{Heinemeyer:2000yj}, which includes
one-loop effects and also the leading two-loop contributions, and gives
results that are somewhat higher than those obtained using~\cite{HHH}. In
order to account for uncertainties in theoretical calculations of $m_h$ in
the MSSM~\cite{Heinemeyer:2000yj} for any given value of $m_t$, we
consider this LEP range~\cite{LEPHiggs} to be consistent with CMSSM
parameter choices that yield $113~{\rm GeV} \le m_h \le 117~{\rm GeV}$.
The theoretical value of $m_h$ in the MSSM is quite sensitive to $m_t$,
the pole mass of the top quark: we use $m_t = 175$~GeV as default, but
mention explicitly the cases where $m_t = 171$~GeV has been used.
Calculations of the Higgs mass and other quantities are also sensitive to
the bottom-quark mass (particularly at large $\tan \beta$), for which we
choose $m_b(m_b)^{\overline {MS}} = 4.25$~GeV for the running mass.  All
but one of the benchmark points we propose satisfy $m_h > 113$~GeV for
$m_t = 175$~GeV.  In view of the expected accuracy $\sim 3$~GeV of the
{\tt FeynHiggs} code, we therefore consider that all the proposed points
are compatible with the LEP lower limit of $113.5$~GeV~\cite{LEPHiggs}.

\subsection{$b \to s \gamma$ Decay}

We implement~\cite{gerri} the new NLO $b \to s \gamma$ calculations
of~\cite{newbsgcalx} when ${\tilde M} > 500$~GeV, where ${\tilde M} = {\rm
Min}(m_{\tilde q}, m_{\tilde g})$. Otherwise, we use only the LO
calculations and assign a larger theoretical error. For the
experimental
value, we combine the CLEO measurement with the recent BELLE
result~\cite{bsgexpt}, assuming full correlation between the experimental
systematics~\footnote{This is conservative, but the available information
does not justify a less conservative approach, and this assumption is in
any case not very important.}, finding ${\cal B} (b \to s \gamma) = (3.21
\pm 0.44 \pm 0.26) \times 10^{-4}$. In our implementation, we allow CMSSM
parameter choices that, after including the theoretical errors 
$\sigma_{th}$ due to the
scale and model dependences, may fall within the 95\% confidence level
range $2.33 \times 10^{-4} < {\cal B}(b \to s \gamma) < 4.15 \times
10^{-4}$. In general, we find in the regions excluded when $\mu < 0$ that
the predicted value of ${\cal B} (b \to s \gamma)$ is larger than this
measured range, whereas, when $\mu > 0$, the exclusion results from ${\cal
B} (b \to s \gamma)$ being smaller than measured.
Table~\ref{tab:derived_quantities} shows the values of ${\cal B}(b \to s
\gamma)$ calculated in our proposed benchmark scenarios.

\begin{table}[htb]
\centering
{\bf Properties of proposed benchmark models}\\
{~}\\
\hspace*{-0.75in}
\begin{tabular}{|c|r|r|r|r|r|r|r|r|r|r|r|r|r|}
\hline
Model          & A   &  B  &  C  &  D  &  E  &  F  &  G  &  H  &  I  &  J  &  K  &  L  &  M   \\ 
\hline
$\ohsq$   & 0.26 & 0.18 & 0.14 & 0.19 & 0.31 & 0.17 & 0.16 & 0.29 & 0.16 &
0.20 & 0.19 & 0.21 &  0.17 \\
\hline
$\delta a_{\mu}$
               & 2.8 & 28 & 13 &-7.4 & 1.7 &0.29 & 27 & 1.7 & 45 & 11 &-3.3 & 31 &  2.1 \\
\hline
$B_{s \gamma}$
           &  $3.54$    &  $2.80$   &  $3.48$  
& $4.07$    &  $3.40$   & $3.32$    & $3.10$    & $3.28$ & $2.55$    & $3.21$   
& $ 3.78$   &  $2.71$   &  $ 3.24$  
\\ 
$\sigma_{th}$ & 0.15 & 0.12 & 0.14 & 0.17 & 0.14 & 0.14 & 0.13 & 0.14 &
0.11 & 0.14 & 0.16 & 0.12 & 0.14 \\
\hline
$\Delta$& 275 &  43 & 108 & 166 &  46 & 325 &  90 &1056 &  76 & 272 & 477 & 128 & 1199  \\ 
(+ $\lambda_t$) & (292) &  (47) & (117) & (177) & (153) & (559) & 
 (97) & (1098) &  (83) & (294) & (537) & (138) & (1276)  \\ 
\hline
$\Delta^\Omega$ &6.0 & 1.3 & 5.7 & 7.0 & 106 & 85 & 9.3 & 36 & 12 & 32 &
91 & 7.3 & 33 \\ 
(+ $\lambda_t$) & (6.0) & (1.3) & (5.9) & (7.0) & (372) & (1089) & (11) &
(36) & (13) & (33) & (125) & (29) & (206) \\ 
\hline
\end{tabular}
\caption[]{\it 
Derived quantities in the benchmark models proposed. In addition to the
relic density $\ohsq$, the supersymmetric
contribution to $a_\mu \equiv (g_\mu - 2)/2$ in units of $10^{-10}$, and
the $b \rightarrow s
\gamma$ decay branching ratio $10^{-4}$, we also display the amount of
electroweak
fine-tuning $\Delta^\Omega$ (all of the above quantities are calculated
using {\tt SSARD}), and the amount of electroweak fine-tuning,
calculated with
the {\tt BMPZ} code \cite{Pierce:1997zz}, using the {\tt ISASUGRA 7.51}
versions of the input parameters.}
\label{tab:derived_quantities}
\end{table}

\subsection{Muon Anomalous Magnetic Moment}

The BNL E821 experiment has recently reported \cite{Brown:2001mg} a new value for the
anomalous magnetic moment of the muon: $g_\mu - 2 \equiv 2 \times a_\mu$, 
which yields an apparent discrepancy with the Standard Model prediction
at the level of 2.6 $\sigma$: 
\begin{equation}
\delta a_\mu \; = \; (43 \pm 16) \times 10^{-10}.
\label{amu}
\end{equation}
The largest contribution to the stated error is due to statistics, and is
expected to be reduced soon by a factor two or more. The systematic errors
reported by the BNL E821 experiment are considerably smaller in magnitude. 
The largest uncertainty in the Standard Model prediction is that due to
the hadronic contributions: $\delta a^{had}_\mu \sim 7 \times 10^{-10}$.
The largest contribution to $a^{had}_\mu$ is in turn due to vacuum
polarization
diagrams, with the most important uncertainty being that in the low-energy
region around the $\rho^0$ peak. The uncertainty in the hadronic vacuum
polarization in this energy region may be reduced by combining the $e^+
e^-$ annihilation data with those from $\tau^\pm \rightarrow \rho^\pm \nu$
decay.
There is also a hadronic contribution from light-by-light scattering
diagrams, which has been estimated using chiral perturbation theory, and
is thought to yield a smaller uncertainty in $a^{had}_\mu$~\cite{CM}. 

The estimate of the hadronic vacuum-polarization
contributions~\cite{Davieretal} used in the E821 paper~\cite{Brown:2001mg}
does not include the latest $e^+ e^-$ data from Novosibirsk~\cite{Novo} and
Beijing~\cite{BES}, nor the most recent $\tau$ decay data from
CLEO~\cite{CLEO}~\footnote{More data on $\tau$ decays can be expected
from the LEP experiments and the $B$ factories.}. However, these are
thought unlikely~\cite{Davierpc}
to change the overall picture:  we recall that the quoted hadronic error
$\sim 7 \times 10^{-10}$ is much smaller than the apparent discrepancy and
the experimental error. Advocates of new physics beyond the Standard Model
may therefore be encouraged.  However, a final conclusion must await the
publication of more $g_\mu - 2$ data and the achievement of consensus on
the hadronic contribution. 

{\it A priori}, the BNL measurement favours new physics at the TeV scale,
and we consider the best motivated candidate to be supersymmetry. Even
before the hierarchy motivation for supersymmetry emerged, the potential
interest of $a_\mu$ was mentioned, and a pilot calculation
performed~\cite{Fayet}, followed by many
others~\cite{GM,LNW}. Some time ago, it was
emphasized~\cite{LNW} that the BNL experiment would be sensitive to a
large range of the parameter space of the CMSSM with universal
soft superymmetry-breaking parameters at the input GUT scale, determining
in particular the sign of the Higgs mixing parameter $\mu$~\cite{LNW}. 
A large number of theoretical papers have discussed the interpretation of
the BNL measurement within supersymmetry~\cite{Feng:2001tr,ENO}. These
calculations generally agree that $\mu > 0$ is favoured by the BNL measurement. 
The calculations we use in this paper are taken from~\cite{ENO}, which are
based on~\cite{IN}~\footnote{For other recent calculations,
see~\cite{Moroi}.}, including also the leading two-loop electroweak
correction factor~\cite{CM2l}.

In this paper, we do not impose the BNL $g_\mu - 2$ constraint in the form
(\ref{amu}), though we do bear it in mind in the selection of points,
for example in the relative weighting of points with $\mu > 0$ and $\mu <
0$.  About half of the points we propose yield values of $\delta a_\mu$
that are compatible with (\ref{amu}) within two standard deviations, and
several of the points lie within one standard deviation.
Table~\ref{tab:derived_quantities} shows the values of $\delta a_\mu$
calculated in our proposed benchmark scenarios.

\subsection{Cosmological Relic Density}

Like most analyses of CMSSM phenomenology for future colliders, we assume
that $R$ parity is conserved. This implies that the lightest
supersymmetric particle (LSP) is stable, and hence should be present in
the Universe today as a cosmological relic from the Big Bang, constituting
part of the dark matter. If the LSP had either strong or electromagnetic
interactions, it would bind with conventional matter to form anomalous
heavy isotopes.  These are not seen down to levels far below the
calculated relic density, so the LSP can have only weak and gravitational
interactions~\cite{EHNOS}. 
There are scenarios in which the LSP is not the supersymmetric partner of
any of the Standard Model particles. For example, it might be the
gravitino or axino. In these cases, cosmological constraints on the dark
matter density cannot be used to constrain the CMSSM in a useful way, and
values of $m_{1/2}$ and $m_0$ larger than those we discuss would also be
allowed. 

Among the supersymmetric partners of Standard Model particles, LEP data
and direct searches for the scattering of cold dark matter particles
appear to exclude the possibility that the LSP is a sneutrino $\tilde \nu$
in the MSSM~\cite{fosi}.  The most viable LSP candidate seems to be the
lightest neutralino $\chi$, and this is the hypothesis adopted here.
Since the sparticle spectrum is explicitly calculable in the
CMSSM, we concentrate on regions of its parameter space in which the LSP
is a neutralino, to the exclusion of other regions. 

Astrophysics and cosmology provide many independent arguments that most of
the gravitating matter in the Universe is invisible. Some of this is
certainly baryonic, but the consistency of cosmological nucleosynthesis 
calculations with the observed light element abundances suggest that most
of the dark matter is 
non-baryonic \cite{bbn}. This conclusion has been reinforced by recent estimates
of the cosmological baryon density $\Omega_b$ based on microwave
background data, which suggest \cite{dasi}
\begin{equation}
\Omega_b h^2 \; \sim \; 0.02,
\label{Omegab}
\end{equation}
with an error of about 20 \%, where $h$ is the present Hubble expansion
rate in units of 100~km/s/Mpc. The Hubble Key Project~\cite{Key} and other
measurements indicate that $h^2 \sim 0.5$, again with an error of
about 20 \%. The estimate (\ref{Omegab}) is much smaller than the
corresponding estimate of the overall matter density \cite{dasi}:
\begin{equation}   
\Omega_m h^2 \; \sim \; 0.14 \pm 0.04.
\label{Omegac}
\end{equation}
We conclude that most of the matter in the Universe is in the form of
non-baryonic dark matter, and hypothesize in this paper that it consists
mainly of the lightest neutralino $\chi$.

For the purpose of this paper, we assume
\begin{equation}   
0.1 \; \le \; \Omega_\chi h^2 \; \le \; 0.3. 
\label{Omegachi}
\end{equation}
The upper limit being a conservative upper bound based 
only on the lower limit to the
age of the Universe of 12 Gyr. Larger values of
$\Omega_\chi h^2$ would require values of $m_{1/2}$ and $m_0$
larger than those
we discuss, in general. Smaller values of $\Omega_\chi h^2$, 
corresponding to smaller
values of $m_{1/2}$ and $m_0$, are certainly possible,
since it is quite possible that some of the cold dark matter might not
consist of LSPs. Axions and ultraheavy metastable relic particles are
other candidates that might contribute. However, allowing smaller values
of $\Omega_\chi h^2$ would open up only a very small extra area of
the $(m_{1/2}, m_0)$ plane, as we see shortly. 

We base our relic density calculations on a recent analysis~\cite{EFGOSi}
using {\tt SSARD} that extends previous results~\cite{EFGO} to larger
$\tan \beta > 20$.  We note here two important effects on the calculation
of $\Omega_\chi h^2$ that were discussed in~\cite{EFGOSi}, which are due
to improvements of previous calculations of $\chi - {\tilde \ell}$
coannihilations and direct-channel $\chi \chi$ annihilations through the
heavier neutral MSSM Higgs bosons $H$ and $A$
\cite{dn,BB,Baer,lns,EFGOSi}. Both of these effects extend the region of
CMSSM parameter space consistent with cosmology out to values of $m_0$ and
$m_{1/2}$ that were larger than those found at smaller values of $\tan
\beta$~\cite{EFOSi,EFGO}. As we discuss later, good overall consistency
was found~\cite{ENO} between these relic density calculations, the LEP and
other sparticle mass limits, the LEP Higgs `signal' and measurements of $b
\to s \gamma$, and also the recent BNL measurement of $g_\mu - 2$ if $\mu
> 0$. There is also a region of the $(m_{1/2}, m_0)$ plane at relatively
large values of $m_0$, close to the higgsino LSP area, termed the
`focus-point' region. This is consistent with $b \to s \gamma$ for any
$\tan\beta$, and may also be consistent with $g_\mu - 2$ if $\tan\beta$ is
large and $m_{1/2}$ is relatively small, according to the {\tt BMPZ}
code \cite{Pierce:1997zz} although not according to {\tt SSARD}
for the input parameter values used in Fig.~\ref{fig:locations2}d (see
the discussion in Sec.~\ref{uncertainties}).

Table~\ref{tab:derived_quantities} shows the values of $\ohsq$
calculated in our proposed benchmark scenarios.

\subsection{Electroweak and Cosmological Fine-Tuning}

Here we discuss two distinct issues: the fine-tuning of CMSSM parameters
that is required to obtain the electroweak scale, and the sensitivity of
the cosmological relic density to input parameters. 

As mentioned in the Introduction, the TeV mass scale for supersymmetry is
largely motivated by the gauge hierarchy problem:
how to make the small electroweak scale $m_Z \ll m_P
\sim 10^{19}$~GeV `natural', without the need to fine-tune parameters at
each order in perturbation theory~\cite{hierarchy}. This is possible if
the supersymmetric partners of the Standard Model particles weigh $\lappeq 1$~TeV,
but the amount of fine-tuning of supersymmetric parameters
required to obtain the electroweak scale increases rapidly for sparticle
masses $\gg 1$~TeV. In an attempt to quantify this, it was
proposed~\cite{EENZ} to consider the logarithmic sensitivities of the
electroweak scale to the supersymmetric model parameters $a_i$:
\begin{equation}
\Delta \equiv \sqrt{ \Sigma_i (\Delta_i)^2}: \Delta_i \equiv
{a_i \over m_Z} {\partial m_Z \over \partial a_i}.   
\label{EENZ}
\end{equation}
In the CMSSM with universal soft supersymmetry-breaking
parameters, the fundamental parameters $a_i$ include the common scalar
mass $m_0$, the common gaugino mass $m_{1/2}$, the common trilinear
parameter $A_0$ at the GUT scale, the supersymmetric Higgs mass parameter
$\mu$ at the GUT scale and the supersymmetry-breaking Higgs mass parameter
$B$ at
the GUT scale. These are the fundamental dimensionful parameters which are
expected to be directly related to the physics responsible for breaking
the electroweak symmetry and generating the correct size for the
electroweak scale.  In view of the sensitivity of the electroweak scale to
the top (and possibly the bottom) Yukawa coupling $\lambda_t (\lambda_b)$,
some (but not all) of the Yukawa couplings at the GUT scale are sometimes
included among the fundamental parameters in (\ref{EENZ})~\footnote{For an
extensive discussion of the philosophy behind these choices, see
~\cite{Feng:2001bp}.}. 
In what follows, we quote the values of $\Delta$ for both cases:
first, considering the fine tuning only with respect to the dimensionful
CMSSM parameters, and then including also the sensitivity to $\lambda_t$.


An analogous measure of the amount of
fine-tuning needed to obtain in the CMSSM a relic density
$\Omega_\chi h^2$ in the range preferred by cosmology has been proposed
recently \cite{EO}:
\begin{equation}
\Delta^\Omega \equiv \sqrt{ \Sigma_i (\Delta^\Omega_i)^2}: \Delta^\Omega_i 
\equiv {a_i \over \Omega_\chi} {\partial \Omega_\chi \over
\partial a_i}.
\label{EO}
\end{equation}
In this case, we hold $m_Z$ fixed, and therefore, the set of 
input parameters $\{a_i\}$ becomes $\{m_0, m_{1/2}, A_0, \tan \beta, sgn(\mu)\}$.
The relic density is also quite sensitive to the values of the
Standard Model parameters $m_t$ and $m_b$.


Table~\ref{tab:derived_quantities} shows the values of $\Delta$ and
$\Delta^\Omega$ calculated in our proposed benchmark scenarios.  The first
(second) row for $\Delta$ shows the electroweak fine-tuning without (with)
$\lambda_t$ included among the $a_i$ (the $\lambda_b$ dependence of
$\Delta$ is relatively mild). The first (second) row for $\Delta^\Omega$
shows the cosmological fine-tuning without (with) $m_t$ and $m_b$ included
among the $a_i$. The dependence of $\Delta^\Omega$ on $m_b$ is
significant at high $\tan \beta$, particularly for point K, L and M. We
see from Table~\ref{tab:derived_quantities} that, as
a rule, the electroweak fine-tuning roughly scales with $m_{1/2}$ and is
independent of $m_0$, if only the sensitivity to the dimensionful
parameters is considered -- this is in essence the focus-point
phenomenon~\cite{Feng:2000hg,Feng:2000mn}. On the other hand,
$\Delta^\Omega$ behaves similarly to $\Delta$, except in the `focus-point'
and rapid-annihilation `funnel' regions, where it is found that there is a
strong sensitivity of $\Omega h^2$ to the input parameters \cite{EO}. 

We emphasize that fine tuning should not be confused with instability: 
the fact that CMSSM model parameters might need to be adjusted carefully
in some cases does not mean that the resulting points in parameter space
are inherently unstable. They are perfectly good electroweak vacua, and
cannot be excluded {\it a priori}. The extent to which one cares about the
amount of fine tuning depends on the underlying measure in CMSSM parameter
space, which is of course unknown at present.  Moreover, there are surely
correlations between the input parameters, and it is known that these may
reduce radically the apparent amounts of fine tuning.  Finally, extending
the spirit of (\ref{EENZ}) and (\ref{EO}), one might define
alternative (or additional) measures of fine-tuning reflecting the
sensitivities of other physical observables to the fundamental parameters,
which might change again the relative weights of the points.  For example,
one could consider large CP-violating phases~\cite{Brhlik:1999zn} and ask
about the degree of fine-tuning required to bring various electric dipole
moments in accord with experiment.  One would then find that in this sense
the bulk points are much more fine-tuned than the focus
points~\cite{Feng:2001bp}, coannihilation `tail' points~\cite{EFOSi} and
rapid-annihilation `funnel' points~\cite{EFGOSi}. For these reasons, measures of
fine tuning come
with impressive health warnings on the packet. Hence, we do not use their
values as selection criteria for benchmarks.  However, we do see clearly
that some models are more finely tuned than others. The values of $\Delta$
vary by a factor of 28 between benchmarks B and M (27
if $\lambda_t$ is included among the $a_i$.), 
whereas the values of $\Delta^\Omega$ vary over
a factor 81 between benchmarks B and E (840 between benchmarks B and F
if $m_t$ is included among the $a_i$). 

\subsection{Combination of Constraints}

The interplays of all these constraints in the $(m_{1/2} , m_0)$ planes
for some values of $\tan \beta$
are illustrated in Fig.~\ref{fig:locations1} and \ref{fig:locations2}.
The very dark (red) triangular regions at large $m_{1/2}$ correspond to
$m_{\sTau_1} < m_{\chi}$, where $\sTau_1$ is the lighter $\tilde
\tau$ mass eigenstate. These regions 
are ruled out by the requirement that the LSP be neutral. We show as (red)
dash-dotted lines the $m_h = 113$~GeV contour calculated using {\tt
FeynHiggs}~\cite{Heinemeyer:2000yj}.
We see that the Higgs mass bound from LEP excludes regions of small
$m_{1/2}$,
and has strongest impact at low $\tan \beta$. In a previous analysis, it
was shown in~\cite{EGNO} that the whole of the plane which is of
cosmological
interest is excluded by the Higgs bound for values of $\tan \beta \lappeq
3.5$. Hence, for the benchmarks, only values of tan$\beta \geq 5$
were considered.
The (dashed) bound on the chargino mass from LEP excludes very low
$m_{1/2}$
values, almost independently of tan$\beta$,  and the LEP
selectron constraint (dot-dashed) excludes a region around the origin in the
$(m_{1/2}, m_0)$ plane. We do not show them on all the panels in
Figs.~\ref{fig:locations1} and \ref{fig:locations2}, but only in panels (a,b)
of 
Fig.\ref{fig:locations1} for $\tan\beta = 5, 10$ and $\mu > 0$: their
locations are similar for the other CMSSM cases studied. The branching
ratio for $b \rightarrow s \gamma$
excludes a dark (green) shaded  area at low $m_{1/2}$.
Its impact increases with increasing tan$\beta$, and
is larger for
$\mu < 0$.

The cosmological constraint $0.1 \leq \Omega_\chi h^2 \leq 0.3$ allows a
region shown in light grey (turquoise), which exhibits a narrowing
coannihilation strip that extends at large $m_{1/2}$ into
the domain where the $\sTau_1$ is the LSP. This defines
upper bounds on the allowed values of $m_{1/2}$ (and hence $m_\chi$)
in the coannihilation region, which are
\begin{equation}
m_{1/2} \; \sim \; 1400~{\rm GeV}, \; m_\chi \; \sim \; 600~{\rm GeV},
\label{cosmoupperlimit}
\end{equation}
for $\tan \beta \lappeq 20$, increasing at larger $\tan \beta$:
$m_{1/2} \la 1900 (2200)$~GeV is allowed for $\tan \beta = 35
(50)$, as seen in (b, c, d) of Fig.~\ref{fig:locations2}. The `tails' of
these regions are potentially beyond the physics reach of the LHC, but are
disfavoured by $g_\mu - 2$.

We also see in panels (c, d) of Fig.~\ref{fig:locations2} the possibility
of a `funnel' extending to large $m_{1/2}$ and $m_0$ where an acceptable
relic density is made possible by rapid direct-channel annihilation $\chi
\chi \rightarrow H, A$. There is also an allowed cosmological strip at
large $m_0$ where the LSP has a significant higgsino component, and as a
result, neutralino annihilation to gauge boson pairs, as well as
$s$-channel Higgs exchange are enhanced \cite{Feng:2000gh,Feng:2001zu}. 
This region lies in the `focus-point' region and is present in all panels
with $\tan\beta>5$, although for improved readability of the figures we
choose to show it only in panel (d)  in Fig.~\ref{fig:locations1}.  Just
above the focus-point region there is a shaded area with no acceptable
electroweak symmetry-breaking solutions, and a light higgsino-like
chargino near its boundary.  Areas in Fig.~\ref{fig:locations1} and
\ref{fig:locations2} between the `focus-point' and the other shaded
cosmological regions have values of $\Omega_\chi h^2$ that are too large,
and hence are excluded by the cosmological relic density constraint. As
already commented, the unshaded areas at lower $m_0$ values have
$\Omega_\chi h^2 < 0.1$, and hence are in principle allowed by
cosmology\footnote{The central regions of the direct $H,A$ annihilation
channels in Figs.~\ref{fig:locations2}(c,d) are also allowed, as the relic
density is very small there. However, the exact position of this region is
sensitive to the input parameters, as the cosmological fine-tuning measure
$\Delta^\Omega$ indicates.}. However, the remaining parts of these regions
compatible with other constraints are quite small~\footnote{We discuss
this point again in Section~6.}.

Finally, we note that the $g_\mu - 2$ result prefers the diagonal band at
low $m_0$ and $m_{1/2}$ shown in darker grey (pink).  The one-sigma band
is indicated by dashed lines and the full lines represent the two-sigma
band.  We see that there is good overall compatibility between $g_\mu - 2$
and the other constraints for $\tan \beta \gappeq 10$ and $\mu > 0$. The
$g_\mu - 2$ constraint disfavours large values of $m_{1/2}$ and $m_0$,
excluding, for example, the tails of the cosmological region. 

\section{Proposed Benchmark Points}

Supersymmetric benchmark points have a venerable history in physics
studies for future
colliders~\cite{ATLASTDR,TESLATDR,Carena:2000yx,epbm}.
They were useful in showing how many spectroscopic measurements
might be possible at the LHC, and in demonstrating the precisions
possible there and with an $e^+ e^-$ linear collider. However, only about
3 out of the 26 points previously studied are clearly
compatible with all the LEP constraints, though some cases may survive if
the theoretical errors in calculating $m_h$ are favourable to them. The
points used previously also could not take into account the recent
constraints from $b \rightarrow s \gamma$ and $g_\mu - 2$. Furthermore,
many of the previous points also give unacceptably large relic densities. 
For instance, six points in CMSSM parameter space were studied in detail
for the LHC~\cite{Paige}. None of them have
survived the most recent
LEP2 limits, dark matter constraints and $g_{\mu} - 2$ constraints. 

We have chosen our proposed new benchmark points for tan$\beta$ = 5, 10,
20, 35 and 50 to span the possibilities in the preferred regions. The
locations of the points in the $(m_{1/2} , m_0)$ planes for different
values of $\tan \beta$ are shown in Figs.~\ref{fig:locations1} and
\ref{fig:locations2}.  As already remarked, the points are probably all
consistent with the LEP Higgs mass constraint $m_h \geq 113.5$ GeV, once
theoretical uncertainties are taken into account. We also took note of the
$g_{\mu} - 2$ measurement, so that most points have $\mu > 0$ and several
are within the 2-$\sigma$ experimental range.  In some cases, points with
different values of tan$\beta$ give rise to very similar particle spectra
and decay characteristics, and it was decided to keep only one example, so
as to avoid duplication.  For this and many other reasons, the chosen
points should not be considered an unbiased statistical sampling of the
CMSSM possibilities. However, we did make an effort to probe the different
possibilities. Thus, we include two `focus-point' models, two in the
coannihilation tails at large $m_{1/2}$, and two in rapid $\chi \chi
\rightarrow H,A$ annihilation funnels, and we kept two points with $\mu <
0$. 

The input parameters of the benchmark points, labelled from A to M,  
are listed in Table~\ref{tab:msugra.spectr_Olive}.

We now make some comments on the individual points.

\begin{itemize}

\item[A]: The only allowed points for this small $\tan \beta = 5$ are far
into the coannihilation tail, and thus have relatively large $m_{1/2}$,
essentially to ensure $m_h \ge 113$~GeV. For this reason, this value of
$\tan \beta$ is now disfavoured by $g_\mu - 2$. It would be possible to
choose a smaller value of $m_{1/2}$ if one made greater allowance for
theoretical error in the $m_h$ calculation, e.g., by choosing $m_t >
175$~GeV and $A_0 > 0$, and relaxing the $g_\mu - 2$ constraint. We note
that, when $\mu < 0$, consistency with the Higgs limit would have required
$m_{1/2} > 830$ GeV, and we do not consider this limited region for
further study.

\item[B]: This point with $\tan \beta = 10$ has much smaller $m_{1/2}$,
and hence $m_h < 113$~GeV in our nominal {\tt FeynHiggs} calculation.
Though it formally fails the Higgs mass bound, it does so just barely:
$m_h = 112$ GeV for this point, which is compatible with
LEP~\cite{LEPHiggs} within the theoretical errors. On the other hand, it
satisfies the $g_{\mu} - 2$ constraint within about one $\sigma$. Thus
points A and B take complementary points of view concerning these two
constraints. 

\item[C]: This second point with $\tan \beta = 10$ is compatible with
$g_{\mu} - 2$ at the 2-$\sigma$ level, as well as with $m_h = 113$~GeV in
our nominal {\tt FeynHiggs} calculation. It is therefore intermediate in
philosophy between points A and B. 

\item[D]: Again with $\tan \beta = 10$, but one of just two points with
$\mu
< 0$. This point has $g_{\mu} - 2$ within about 3 $\sigma$ of the current
experimental value. Like point C, it is compatible with $m_h = 113$ GeV
according to {\tt FeynHiggs}.  Note that the effect of the $b \to s \gamma$
limit is similar to that of the Higgs limit over the range of $m_0$ favoured by
the relic density. Both constraints force this point into the coannihilation
region.

\item[E]: The first of two `focus-point' models with large $m_0$ at $\tan
\beta = 10$. Note that {\tt SSARD} uses $m_t = 171$ GeV for this and the
other focus point, whereas the {\tt ISASUGRA 7.51} version uses $m_t =
175$~GeV. For this reason, {\tt FeynHiggs} yields somewhat different
values of $m_h = 112, 116$~GeV, respectively. Both of these are compatible
with the LEP lower limit, taking into account theoretical uncertainties. 
With {\tt SSARD}, focus-point solutions could also be obtained with the
default choice of $m_t = 175$~GeV, but at larger $m_0$ for the same
value of $m_{1/2}$. 

\item[F]: This second focus-point model has larger $m_{1/2}$ and $m_0$,
and so has a higher Higgs mass: $m_h = 115, 121$~GeV for the {\tt SSARD}
and {\tt ISASUGRA 7.51} versions. Like point E, this point uses $m_t =
171$ GeV. Both points are about 2.5 $\sigma$ away from the central value
of $g_{\mu} - 2$. We note that the focus-point region extends to larger
$m_{1/2}$ (and $m_0$).

\item[G]: One of two points with moderate $\tan \beta = 20$, this one is 
chosen to have the relatively low value $m_h = 114$ GeV. It is also
just consistent with the $b \rightarrow s \gamma$ constraint, and agrees
with the $g_{\mu} - 2$ measurement at the 1-$\sigma$ level.

\item[H]: A second point with the same moderate $\tan \beta = 20$, but
this time at the end point of the coannihilation `tail' with very large
$m_{1/2}$.  Correspondingly, it has a larger Higgs mass: $m_h = 121$~GeV, and a
small value of $g_{\mu} - 2$, about 2.5 $\sigma$ from the central experimental
value. The small $m_{\tilde \tau_1} - m_\chi$ mass difference has its own
interesting features and challenges, as we discuss later. 

\item[I]: A very ($g_{\mu} - 2$)-friendly point at large $\tan \beta =
35$. Note that the inclusion of one-loop corrections to $m_\chi$ are
essential here. If they are not included, this point has ${\tilde \tau_1}$
as the LSP. The $b \rightarrow s \gamma$ constraint is the dominant one at
small $m_{1/2}$ for this value of $\tan\beta$.

\item[J]: A second point with $\tan \beta = 35$ and $\mu > 0$, this time
about half-way along the coannihilation `tail' at large $m_{1/2}$, that is
compatible with $g_{\mu} - 2$ at the 2-$\sigma$ level. 

\item[K]: One of two points in a rapid $\chi \chi \rightarrow H,A$
annihilation `funnel', and one of just two points with $\mu < 0$.  This
point has $\tan \beta = 35$, which is (almost) the largest value where we
find consistent electroweak vacua for this sign of $\mu$, with our default
choices of the auxiliary parameters $m_t, m_b$ and $A_0$. It is far from
saturating the experimental constraints, apart from $g_\mu - 2$. We note
that the rapid-annihilation `funnel' also extends to larger $m_{1/2}$ (and
$m_0$). 

\item[L]: One of two points with (almost) the largest value of $\tan \beta
= 50$ for which we find consistent electroweak vacua for $\mu > 0$. This
point lies in the ($g_{\mu} - 2$)-friendly `bulk' of the cosmological
region, and is highly compatible with the other experimental constraints. 

\item[M]: A second point in a rapid $\chi \chi \rightarrow H,A$
annihilation `funnel', again for (almost) the largest value of $\tan \beta
= 50$ allowed for $\mu > 0$ with our default choices of the auxiliary
parameters. This point has small $g_{\mu} - 2$, but satisfies all the
other experimental constraints.

\end{itemize}

It is characteristic of all the solutions that the lightest Higgs strongly
resembles a Standard Model Higgs boson, whilst the other Higgses are heavy
and nearly degenerate in mass. The LSP is in all cases almost a pure 
${\tilde B}$, except in the focus-point cases, where a non-negligible
${\tilde H}$ component is also present.

\section{Theoretical Uncertainties and Comparisons between Codes}
\label{uncertainties}

Our proposed benchmark points were chosen using the code {\tt
SSARD}~\cite{SSARD} to run all the input parameters down to the
electroweak scale, impose the electroweak symmetry-breaking conditions
there, and evaluate the physical chargino and neutralino masses including
one-loop radiative corrections, which are generally ${\cal O}(5)$\%. This
code does not include the one-loop corrections to slepton masses, which
are generally ${\cal O}(1)$\%, taking values that increase with $\tan
\beta$ and decrease with $m_{{\tilde \tau}_1}$~\cite{Pierce:1997zz}. As
mentioned earlier, the {\tt FeynHiggs} code~\cite{Heinemeyer:2000yj} is
used to evaluate the Higgs spectra.

\begin{figure}[ht]
\begin{center}
\hspace*{.20in}
\begin{minipage}{8.0in}
\epsfig{file=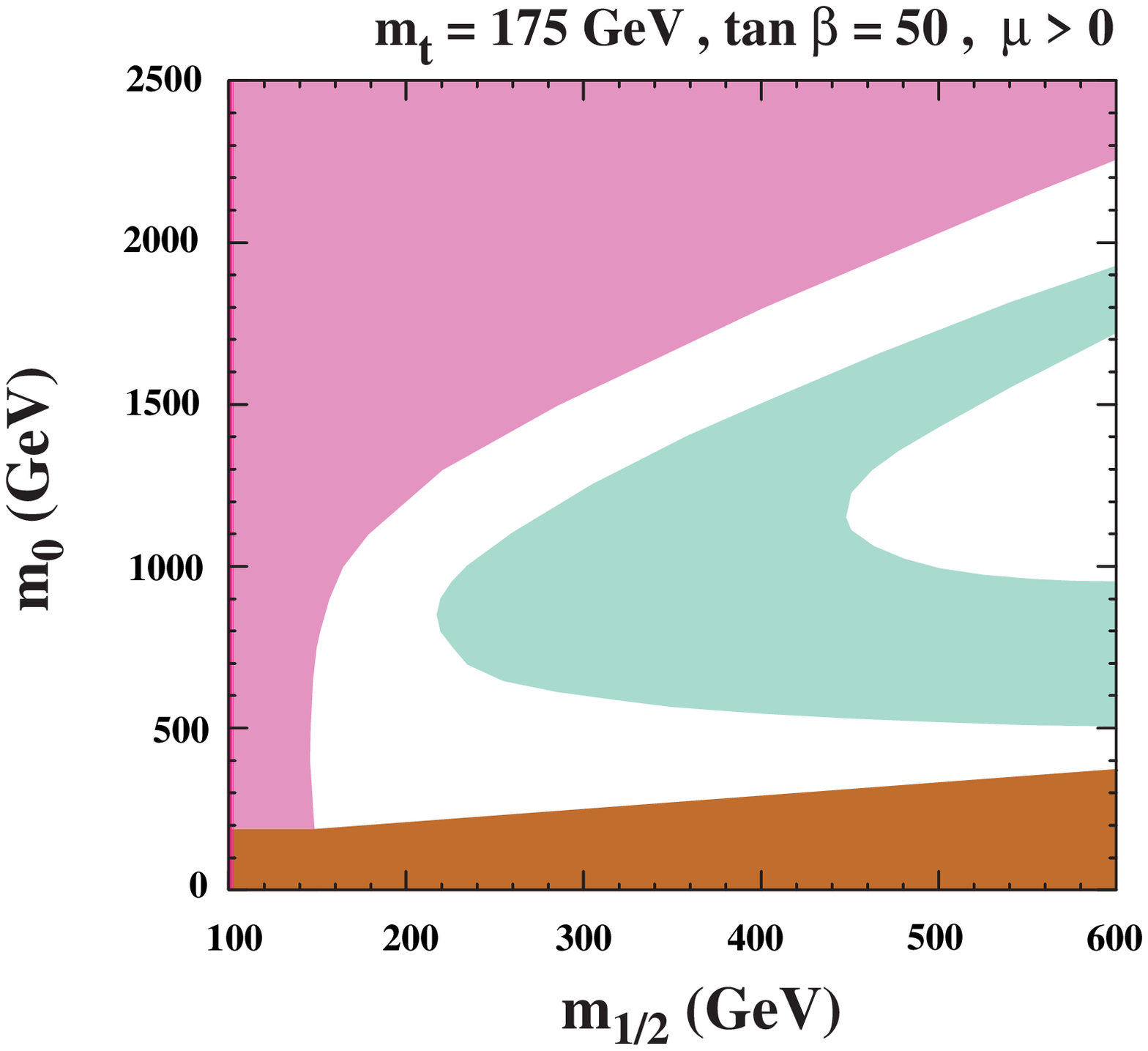,height=2.65in, width=2.6in}
\hspace*{0.5in}
{\vspace*{-0.1in}
\epsfig{file=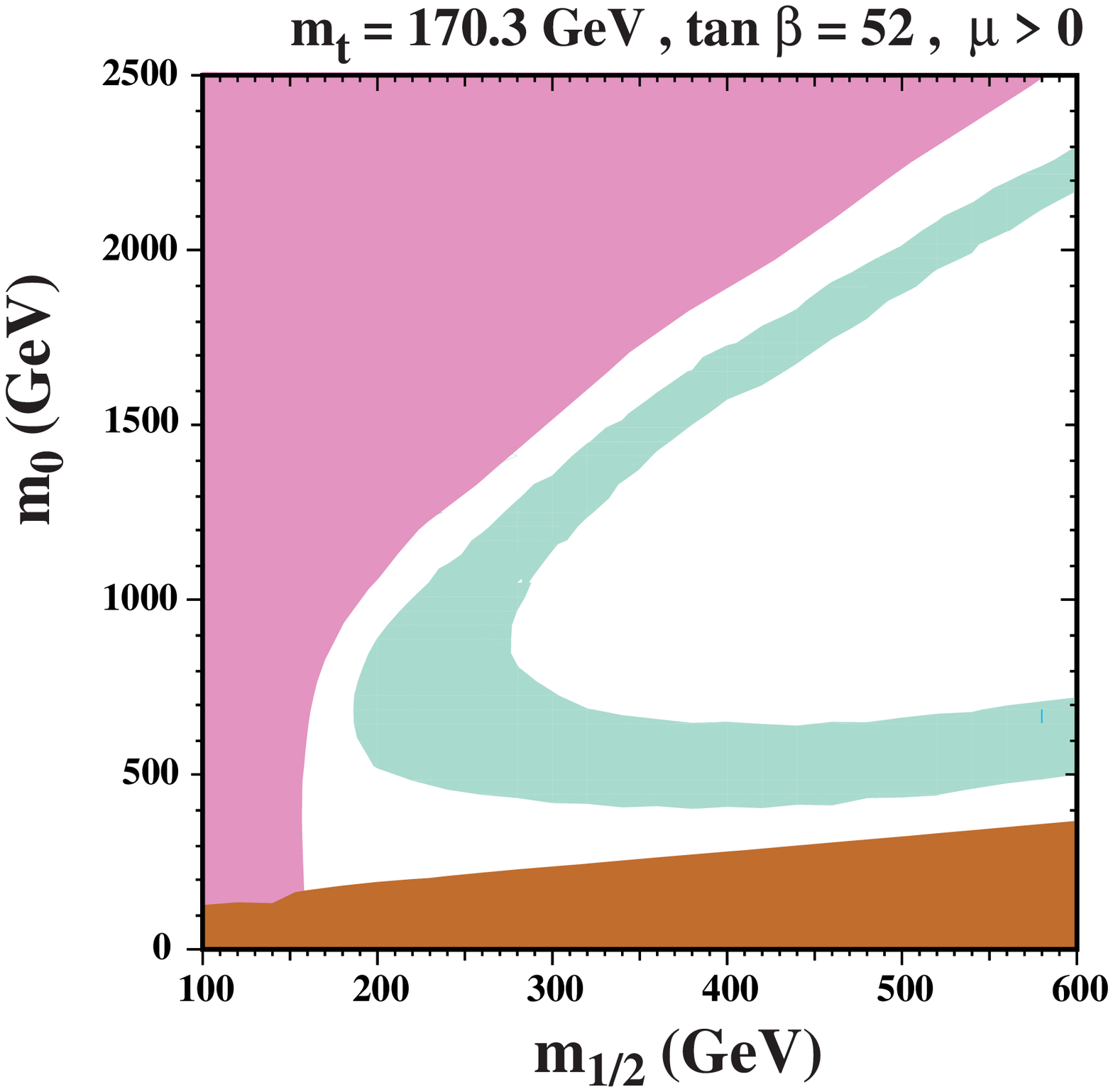,height=2.5in}} 
\end{minipage}
\caption{\label{fig:comparison}   
\it Left: the cosmologically preferred region obtained with {\tt
Neutdriver}~\cite{neut} and mass spectra from the {\tt BMPZ}
code~\cite{Pierce:1997zz},
for the same parameters as in Fig.~\ref{fig:locations2}d.
Right: the corresponding result from {\tt SSARD}, but 
for $\tan\beta=52$ and $m_t=170.3$~GeV.}
\end{center}
\end{figure}

Some words of caution are in order. There are several available programs
for calculating the supersymmetric particle spectrum and computing the
physical observables considered in this paper. In general, these programs
use different algorithms and, depending on their purpose, may have
different levels of sophistication. The magnitudes of the resulting
differences depend on the particular observable considered, on whether
the results are expressed as functions of GUT-scale input parameters or
physical masses, etc.. For example, in Fig.~\ref{fig:comparison}a we show
the cosmologically preferred region \cite{Feng:2000gh} which was
obtained with {\tt Neutdriver} \cite{neut} 
and mass spectra from the {\tt BMPZ} code \cite{Pierce:1997zz},
for the same parameters as in Fig.~\ref{fig:locations2}d.
We can see that the relic densities found for the same values of $m_{1/2}$
and $m_0$ can be quite different,
reflecting in part the sensitivity $\Delta^\Omega$
(\ref{EO})~\cite{EO} noted earlier, which is generally larger at large
$\tan \beta$~\footnote{The results of~\cite{Baer} are not so dissimilar
from those of~\cite{EFGOSi}. We are aware of other calculations underway
(G.~B\'elanger, F.~Boudjema, A.~Djouadi, M.~Drees, A. Lahanas and
L.~Roszkowski, private communications), which also give rather diverse answers.}. 
In Figs.~\ref{fig:locations2}d and \ref{fig:comparison}a, there are points at
intermediate values of $m_0$ and $m_{1/2}$ where the 
results for $\ohsq$ can differ by as much as a factor of 20.
We have sought to understand the origin of this
apparent discrepancy and the related theoretical uncertainty
in the calculations.

Comparing codes, we find that the bulk of the effect is due to differences
in the supersymmetric mass spectrum, most notably the values for the CP-odd
Higgs mass $m_A$ and the $\mu$ parameter.  We remind the reader that
although both {\tt SSARD} and {\tt BMPZ} are NLO programs, numerical
differences do arise at NNLO due to a different treatment of the NNLO
terms. We list some of these effects below:

\begin{itemize}

\item The treatment of gauge coupling unification. In {\tt SSARD}, all
three gauge couplings are unified at the GUT scale, and the resulting
prediction for the weak scale value of $\alpha_s$ is shown in
Table~\ref{tab:msugra.spectr_Olive}.  On the other hand, in {\tt BMPZ}
({\tt ISASUGRA}), $\alpha_s$ is treated as an input at the weak scale, with
the following values: $\alpha_s=0.119$ (0.118). The three gauge couplings
do not in general unify, and model-dependent GUT-scale threshold
corrections are assumed to account for the mismatch. This difference in the
value of $\alpha_s$ affects the evolution of $\lambda_t$ between $m_Z$ and
$m_{GUT}$, as well as the extraction of $\lambda_b$ from $m_b(m_b)$.

\item Differences in the extraction of $\lambda_t(m_Z)$, which amount to a
$\sim2.5\%$ effect. Most of this uncertainty comes from using the running
top quark mass ({\tt SSARD})  versus the pole mass ({\tt BMPZ}) as an
argument in the one-loop correction to $m_t$~\footnote{Both {\tt BMPZ} and
{\tt SSARD} include the SM two-loop $\overline{\rm MS}$ contribution
\cite{Gray:1990yh}, assuming it is close to the $\overline{\rm DR}$
value.}.  Given that the one-loop correction to $m_t$ is typically about
11\%, the resulting differences in $\lambda_t$ are well within the NNLO
uncertainty, but have a major impact (up to $\sim 30\%$ at large $m_0$) on
the extracted
value of the $\mu$ parameter. 

\item Imposing electroweak symmetry breaking. One can choose to minimize
the effective potential at the scale $m_Z$ ({\tt SSARD}), or at the scale
of the average stop mass $Q=\{m_{\tilde t_1}m_{\tilde t_2}\}^{1/2}$ ({\tt
BMPZ and ISASUGRA}). In the former case, care is taken to include
corrections ${\cal O}({\rm ln}(Q/m_Z))$, in order to avoid spurious
differences in the values of the $\mu$ parameter, which affects the
gaugino-higgsino mixing and the pseudoscalar mass $m_A$. The latter has a
major effect in regions where $s$-channel
annihilation through $A$ exchange is dominant. 

\item Computation of the Higgs sector. All of the effects already mentioned
have an impact on the calculation of the Higgs boson masses.  Furthermore,
{\tt SSARD} uses the results of \cite{Carena:2000yi}, which include some
two-loop effects, while {\tt BMPZ and ISASUGRA} apply the one-loop Higgs
mass corrections \cite{Pierce:1997zz} at the scale $Q$. The $m_A$
calculation is also affected by numerical differences in the extracted
values of $\lambda_b$. Altogether, we find that the values for $m_A$ in the
two cases can differ by as much as $\sim 20\%$. 

\end{itemize}

Differences of such magnitudes are usually not important
for collider phenomenology. However, this is not the case
with the relic density calculation at large values of $\tan\beta$, because 
of the high sensitivity (\ref{EO}).
Recall that then the dominant annihilation channel is $s$-channel
$A$ exchange into bottoms,
and that the annihilation cross section scales as
\begin{equation}
\sigma \ \sim \
{m_\chi^2 m_b^2 \over m_W^2 (4m_{\chi}^2-m_A^2)^2}\ N_1^2 N_3^2
\tan^2 \beta,
\end{equation}
where $N_{1,3}$ are the Bino and Higgsino components of the LSP.
The product $N_1 N_3$ is relatively large when
$(|\mu| - M_1)^2 \la M_Z^2$.
It is easy to see that in regions of parameter space where one is
not very far away from the $A$ pole, and $\mu$ is not very large,
the $\sim20\%$ effects in $m_A$ and $\mu$ can combine to an $\sim10$
effect in the $\ohsq$ calculation.

It is possible to vary slightly
the input parameters and obtain reasonable agreement for the spectra
obtained using {\tt SSARD} and {\tt BMPZ}/{\tt ISASUGRA 7.51}.
Panel (b) of Fig.~\ref{fig:comparison}b shows the cosmologically preferred
dark matter region obtained with {\tt SSARD} using slightly different
inputs: $\tan\beta=52$ and $m_t=170.3$~GeV.  We now see improved
agreement (to within a factor of 2) between Figs.~\ref{fig:comparison}a and b.
We have isolated several factors that contribute to this remaining difference,
all of them related to the dark matter calculation. One is the treatment of
the bottom-quark radiative corrections and bottom-quark mass, whose
treatment in {\tt SSARD} was discussed in~\cite{EFGOSi}. {\tt Neutdriver}
typically uses a higher value of $m_b$, which leads to a lower relic
density. {\tt Neutdriver} also does not integrate the Boltzmann equations,
but uses an analytic approximation~\cite{EHNOS} based on non-relativistic
expansions of the annihilation cross sections. Other code differences
concern the extent to which coannihilation effects are included, and the
treatment of $s$-channel annihilation rates.  Therefore, in comparing
results over the CMSSM parameter plane, one must be sure not only to
understand the differences in the spectrum codes, but also those in the
codes used to calculate the relic density. 

The above discussion has focused on CMSSM spectra produced by two fully NLO
codes, and we have seen that in spite of the present theoretical
uncertainties, relatively well-matched spectra can be obtained, e.g., by
varying the input parameters. Although very versatile, the codes
generally available
for Monte Carlo simulations of supersymmetry, such as {\tt
ISASUGRA}~\cite{isasugra}, {\tt SPYTHIA}~\cite{Mrenna:1997hu} and {\tt
SUSYGEN}~\cite{susygen}, do not always include all of the ingredients of a
complete NLO analysis. For example, all three programs so far lack the
one-loop radiative corrections to the chargino and neutralino
masses~\footnote{These
are important not only for relating physical masses to GUT-scale input
parameters, but also for relating searches for sparticle species, e.g.,
$e^+ e^- \rightarrow \chi^+ \chi^-$ and $\chiz_1 \chiz_2$.}. For
the convenience of the experimental simulations that frequently use one of
these codes, we have made searches in the input parameter space of {\tt
ISASUGRA 7.51} to find points that reproduce specific features of the
spectra of the different models shown in Table~1. In some cases,
significant differences are inevitable, and compromises have been made. The
non-implementation of radiative corrections to the chargino and neutralino
masses is a particular snag, and we hope that this can be overcome in
future issues of {\tt ISASUGRA}, {\tt SPYTHIA} and {\tt SUSYGEN}. The {\tt
ISASUGRA} parameters that best reproduce the most relevant features of the
spectra of Table~\ref{tab:msugra.spectr_Olive} are shown in
Table~\ref{tab:msugra.spectr_isasugra}, and we use them for the discussion
of decay signatures in this Section. Despite these differences, we note
that the general agreement between our calculated spectra and those
generated by {\tt ISASUGRA 7.51} is good\footnote{For example, for the {\em
same} input parameters, the percent difference is generally not larger than
2\% across the sparticle spectrum. The biggest differences (10-20\%) are
found for the heavy Higgs masses for points K, L and M.  }. The most severe
differences occur in the focus-point region and at large $m_{1/2},m_0$ at
large $\tan \beta$. We note in passing some of the problems that arise when
trying to match {\tt ISASUGRA 7.51} spectra. 

For this purpose, it is convenient to separate the proposed benchmarks
into three classes of points.

\begin{itemize}

\item Points in the `bulk' and the coannihilation `tail'.  Here the most
relevant masses to fit are those of the lightest neutralino $\chi$ and the
lightest
stau ${\tilde \tau}_1$, since the other particles are less relevant for
the dark matter calculation. It is easy to fit $m_\chi$ and $m_{{\tilde
\tau}_1}$ by varying just $m_{1/2}$ and $m_0$: one first varies $m_{1/2}$
to fit the neutralino mass and then adjusts $m_0$ to fit the stau mass. In
these cases, the most relevant fact is that {\tt ISASUGRA 7.51} does not
include the one-loop corrections to the neutralino and chargino masses,
and hence returns lower neutralino masses for the same input values of
$m_0$ and $m_{1/2}$. In order to compensate for this, one has to crank
$m_{1/2}$ up in order that the tree-level mass from {\tt ISASUGRA 7.51}
looks like the one-loop-corrected mass. However, the stop masses also
depend strongly on $m_{1/2}$, and therefore increase when this is done, as
can be seen by comparing Tables~\ref{tab:msugra.spectr_Olive} and
~\ref{tab:msugra.spectr_isasugra}. In turn, once the stop masses are
higher, the Higgs mass also increases, as also seen in
Table~\ref{tab:msugra.spectr_isasugra}. It would, in principle, be
possible also to vary the other input parameters so as to improve the
match, but there is no well-defined procedure for doing this, and it is
not clear what would be learned from such a lengthy exercise. Since the
rest of the spectrum has intrinsic uncertainties anyway, we have not
striven for perfect matches~\footnote{We note in passing that it was
necessary to fix a minor glitch in {\tt ISASUGRA} 7.51 in order to find
a solution for point H. We are grateful to I.~Hinchliffe and F.~Paige
for their help.}. 

\item Focus points. These points are basically defined by the LSP mass and
a certain value for the gaugino-higgsino mixing, which is in turn
determined by the ratio of the one-loop corrected values of the $U(1)$
gaugino mass $M_1$, and $\mu$. Here, the absence of one-loop corrections
in {\tt ISASUGRA 7.51} is again unfortunate.  In these cases, we have
tried to match the masses of the LSP and $\chi^0_2$, since they correspond
to similar values of the effective $M_1$ and $\mu$ values.  Here again, we
vary only a couple of parameters: $m_{1/2}$ and $m_0$, which determines
$\mu$ and therefore the $\chi^0_2$ mass. The change in $m_0$ alters the
slepton masses by about 50~GeV, as seen by comparing
Tables~\ref{tab:msugra.spectr_Olive} and
~\ref{tab:msugra.spectr_isasugra}. 

\item Points in the rapid $\chi \chi \to A,H$ annihilation region.
In these cases, the relevant masses are $m_{1/2}$ and $m_A$. The first
can still be fit, but $m_A$ is only a very slowly varying function of
$m_0$. Since $m_A$ varies more rapidly with $\tan\beta$, we vary this
parameter instead. For convenience, since we use a lower value of $\tan
\beta $ for point M, we use it for point L as well.

\end{itemize} 

\begin{table}[p!]
\centering
\renewcommand{\arraystretch}{0.95}
{\bf Supersymmetric spectra calculated using {\tt ISASUGRA 7.51}}\\
{~}\\
\begin{tabular}{|c||r|r|r|r|r|r|r|r|r|r|r|r|r|}
\hline
Model          & A   &  B  &  C  &  D  &  E  &  F  &  G  &  H  &  I  &  J  &  K  &  L  &  M   \\ 
\hline         
$m_{1/2}$      & 613 & 255 & 408 & 538 & 312 & 1043& 383 & 1537& 358 & 767 & 1181& 462 & 1953 \\
$m_0$          & 143 & 102 &  93 & 126 & 1425& 2877& 125 & 430 & 188 & 315 & 1000& 326 & 1500 \\
$\tan{\beta}$  & 5   & 10  & 10  & 10  & 10  & 10  & 20  & 20  & 35  & 35  & 39.6& 45  & 45.6 \\
sign($\mu$)    & $+$ & $+$ & $+$ & $-$ & $+$ & $+$ & $+$ & $+$ & $+$ & $+$ & $-$ & $+$ & $+$  \\ 
$A_0$          & 0   & 0   &  0  &  0  &  0  &  0  &  0  &  0  & 0   & 0   & 0   &  0  &  0   \\
$m_t$          & 175 & 175 & 175 & 175 & 175 & 175 & 175 & 175 & 175 & 175 & 175 & 175 & 175  \\ 
\hline
Masses         &     &     &     &     &     &     &     &     &     &     &     &     &      \\ 
\hline
$|\mu(Q) |$    & 768 & 343 & 520 & 662 & 255 & 548 & 485 &1597 & 454 & 876 &1213 & 560 & 1842 \\ 
\hline
h$^0$          & 116 & 113 & 117 & 117 & 116 & 121 & 117 & 124 & 117 & 121 & 123 & 118 &  125 \\
H$^0$          & 893 & 387 & 584 & 750 &1435 &2955 & 521 &1813 & 431 & 851 &1070 & 472 & 1737 \\
A$^0$          & 891 & 386 & 583 & 749 &1434 &2953 & 521 &1812 & 430 & 851 &1069 & 471 & 1735 \\
H$^{\pm}$      & 895 & 394 & 589 & 754 &1437 &2956 & 527 &1815 & 440 & 856 &1074 & 481 & 1739 \\ 
\hline
$\chi^0_1$     & 252 &  98 & 164 & 221 & 119 & 434 & 154 & 664 & 143 & 321 & 506 & 188 &  854 \\
$\chi^0_2$     & 467 & 179 & 303 & 414 & 197 & 546 & 285 &1217 & 265 & 594 & 932 & 349 & 1558 \\
$\chi^0_3$     & 770 & 349 & 524 & 667 & 262 & 551 & 491 &1599 & 460 & 879 &1215 & 564 & 1843 \\
$\chi^0_4$     & 785 & 370 & 540 & 674 & 317 & 845 & 506 &1608 & 475 & 889 &1225 & 578 & 1855 \\
$\chi^{\pm}_1$ & 467 & 179 & 303 & 414 & 193 & 537 & 285 &1217 & 265 & 594 & 932 & 349 & 1558 \\
$\chi^{\pm}_2$ & 784 & 370 & 540 & 676 & 317 & 845 & 506 &1608 & 476 & 890 &1225 & 579 & 1855 \\ 
\hline
$\tilde{g}$    &1357 & 606 & 932 &1203 & 804 &2372 & 880 &3186 & 828 &1669 &2516 &1051 & 4029 \\ 
\hline
$e_L$, $\mu_L$ & 435 & 206 & 293 & 383 &1433 &2942 & 290 &1092 & 308 & 599 &1260 & 450 & 1957 \\
$e_R$, $\mu_R$ & 271 & 145 & 182 & 239 &1427 &2897 & 194 & 709 & 234 & 425 &1088 & 370 & 1658 \\
$\nu_e$, $\nu_{\mu}$
               & 428 & 190 & 282 & 375 &1431 &2941 & 278 &1089 & 298 & 593 &1258 & 443 & 1955 \\
$\tau_1$       & 269 & 137 & 175 & 233 &1415 &2873 & 166 & 664 & 159 & 334 & 931 & 242 & 1249 \\
$\tau_2$       & 435 & 209 & 295 & 384 &1427 &2930 & 296 &1081 & 319 & 589 &1204 & 439 & 1809 \\ 
$\nu_{\tau}$   & 428 & 189 & 281 & 374 &1425 &2929 & 275 &1076 & 285 & 571 &1197 & 409 & 1803 \\  
\hline
$u_L$, $c_L$   &1211 & 546 & 833 &1075 &1519 &3397 & 789 &2834 & 756 &1508 &2398 & 978 & 3789 \\
$u_R$, $c_R$   &1167 & 529 & 803 &1036 &1515 &3360 & 764 &2716 & 732 &1452 &2315 & 948 & 3643 \\
$d_L$, $s_L$   &1214 & 552 & 837 &1078 &1521 &3398 & 793 &2835 & 760 &1510 &2400 & 982 & 3790 \\
$d_R$, $s_R$   &1161 & 531 & 801 &1032 &1515 &3356 & 762 &2703 & 730 &1445 &2305 & 945 & 3631 \\
$t_1$          & 940 & 400 & 635 & 845 & 987 &2401 & 601 &2288 & 569 &1190 &1883 & 744 & 3016 \\
$t_2$          &1172 & 580 & 830 &1039 &1292 &2967 & 785 &2649 & 742 &1405 &2122 & 918 & 3378 \\ 
$b_1$          &1126 & 503 & 769 & 998 &1281 &2961 & 713 &2619 & 647 &1335 &2053 & 819 & 3308 \\
$b_2$          &1161 & 534 & 803 &1028 &1503 &3333 & 762 &2667 & 725 &1406 &2121 & 913 & 3388 \\ 
\hline
\end{tabular}
\caption{\it
Mass spectra in GeV for CMSSM models calculated with {\tt ISASUGRA 7.51}.
The renormalization-group equations for the couplings and the soft
superymmetry-breaking parameters include two-loop effects, and the
dominant one-loop supersymmetric threshold corrections to the third
generation Yukawa couplings are included. 
The Higgs potential is minimized at the scale 
$Q=(m_{\tilde t_1}m_{\tilde t_2})^{1/2}$. 
The Higgs and gluino masses are calculated at one loop.
The rest of the superpartner spectrum is calculated at tree level
at the scale $Q$.
The input parameters have been adjusted so that the spectra best
approximate those shown in Table~\ref{tab:msugra.spectr_Olive}.  
We have used the {\tt ISASUGRA 7.51} default values $m_b^{pole}=5$~GeV
and $\alpha_s(m_Z)=0.118$.
It is assumed that $A_0 = 0$, $m_t = 175$~GeV.
}
\label{tab:msugra.spectr_isasugra}
\end{table}

At the end of this lengthy discussion, it
should be clear that the benchmark points are defined by the physical
spectrum and not the values of $m_0$, $m_{1/2}$, $\tan\beta$ etc., which
are attached to them. The latter are nothing but convenient labels, which
may vary from one program to another. In view of its versatility, in the
rest of this paper, we use {\tt ISASUGRA} to discuss sparticle decays and
experimental signatures. The physics, of course, should remain unchanged,
as long as the physical spectrum is the same.  We stress that most of the
differences discussed above are higher-order effects and represent in part
the theoretical uncertainty in the calculation of the sparticle spectrum.


\section{Decay Branching Ratios}

\begin{figure}[htb]
\begin{center}
\resizebox{12cm}{!}{\includegraphics*[0,0][545,360]{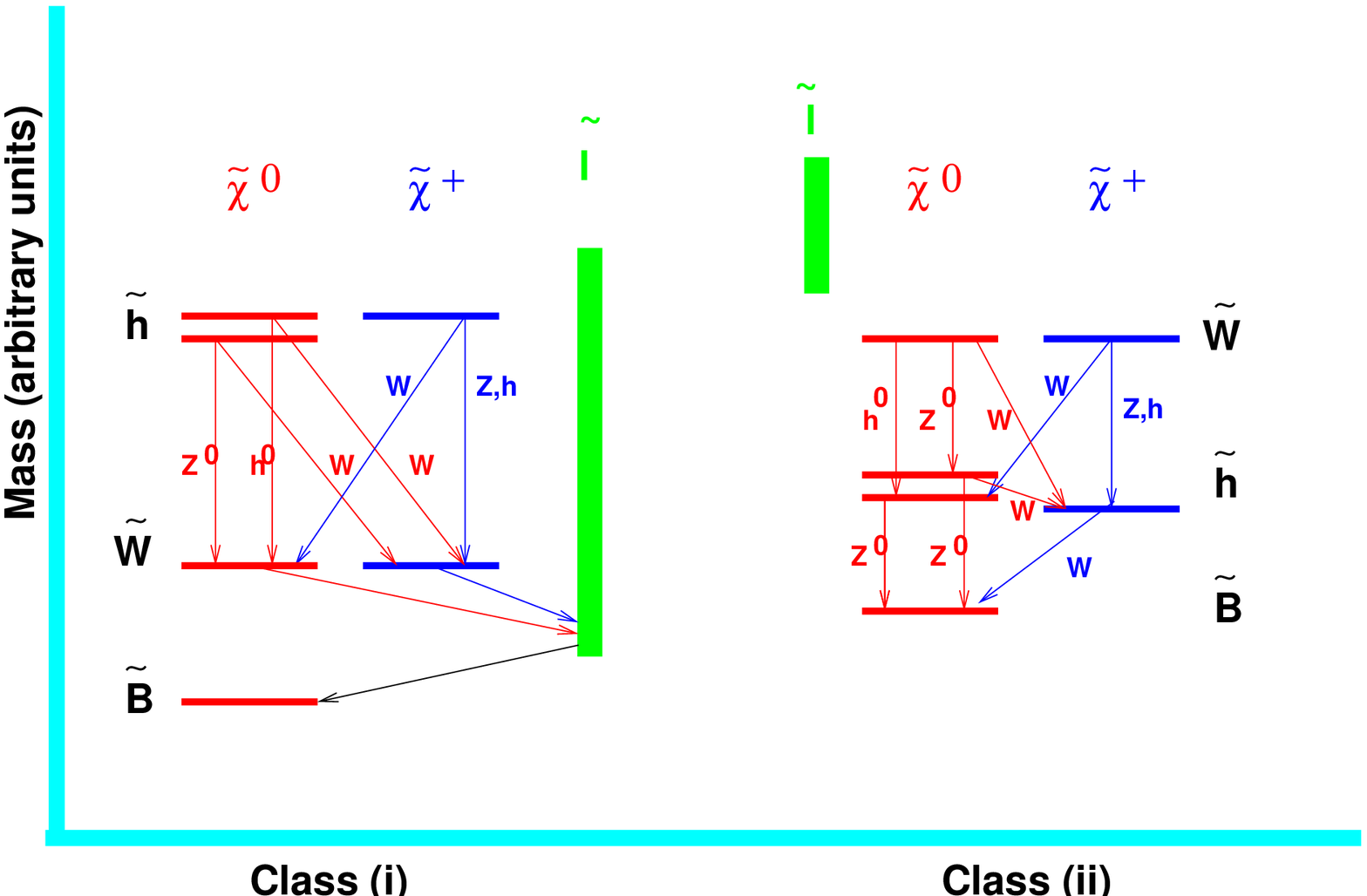}}
\caption{\it Characteristic features of the spectra and principal decay
modes in the two classes of benchmark points.}
\label{fig:features}
\end{center}
\end{figure}

The decay branching ratios in all the proposed scenarios have been
evaluated using the program {\tt ISASUGRA 7.51}.  The benchmark points can
be subdivided roughly into two classes, distinguished by the organization
of their particle masses and their dominant decay modes.  The most salient
features of the decay signatures are displayed in Fig.~\ref{fig:features},
and can be summarized as follows.

(i) For points A-D and G-M, we find that $| \mu | > M_2$, which has the
following consequences:

\begin{itemize}

\item The $\chiz_i$ mass sequence starts as
$\tilde{B}^0$, $\tilde{W}^0$ with some admixture of
$\tilde{H}_{1,2}^0$.  The remaining two states are mainly a
combination of the two Higgsinos with some admixture of 
$\tilde{W}^0$, a pattern typical of the gaugino region.
Furthermore, $m_0 \leq m_{1/2}$, giving
at least one slepton lighter than the $\chiz_2$ and $\chipm_1$, 
and frequently several.

\item The $\chiz_2$ decays mostly to $\sLep \ell$ and $\sNu \nu$ and the
$\chipm_1$ to $\sLep \nu$ and $\sNu \ell$.  These decays are followed in
most cases by $\sLep \rightarrow \ell \chiz_1$.

\item The $\chiz_3$ and $\chiz_4$ decay in $50-60\%$ of the cases into
$\chipm_1 + W^{\mp}$. Other principal decays are about $20-30\%$ $\chiz_3
\rightarrow \chiz_2 Z^0$ and about $15-25\%$ $\chiz_4 \rightarrow \chiz_2
h^0$. 

\item The $\chipm_2$ decays about equally into $\chiz_2 + W^{\pm}$,
$\chipm_1 + Z^0$ and $\chipm_1 + h^0$. 

\item The squarks are in all cases heavier than any of the gauginos or sleptons
and gluinos are heavier than squarks.  They usually have large branching
ratios for cascade decays.  
\end{itemize}
These benchmark points are therefore
characterized by a large proportion of final states with leptons (plus
jets).

(ii) For points E and F, corresponding to the focus-point scenario, we
find $M_1 < | \mu | < M_2$, with the following consequences: 

\begin{itemize}

\item The neutralino mass sequence is roughly:  $\chi \sim \tilde{B}^0$,
$\chiz_2 \sim \tilde{h}^0$, $\chiz_3 \sim \tilde{h}^0$, $\chiz_4 \sim
\tilde{W}^0$, but with large mixing among the states. As $m_0 \gg
m_{1/2}$, all sfermions are considerably heavier than the gauginos. 

\item The dominant neutralino decays are $\chiz_{2,3} \rightarrow \chiz_1
+
Z^0$, $\chiz_{3,4} \rightarrow\chipm_1 + W^{\mp}$, and to a smaller extent 
$\chiz_4 \rightarrow \chiz_2 + Z^0$ or $\chiz_3 + h^0$. 

\item For chargino decays, we find $\chipm_1 \rightarrow \chiz_1 +
W^{\pm}$ with essentially a 100\% branching ratio, and $\chipm_2
\rightarrow \chiz_{2,3} + W^{\pm}$ or $\chipm_1 + Z^0$ or $\chipm_1 +
h^0$.

\end{itemize} The dominance of decays involving $W^{\pm}$ or $Z^0$ leads
to final states with mainly jets and rarely leptons.  These points all
have a gluino lighter than the squarks. 

The above summary lists only the gross features of the benchmark points
and, within the two classes of models, the individual points show
important differences and will sometimes (e.g., solution K)  deviate in
significant details from the above statements.  Taken together, the
benchmarks cover a large variety of cases. 

\begin{figure}[p]
\begin{center}
\resizebox{8cm}{!}{\includegraphics*[0,0][545,800]{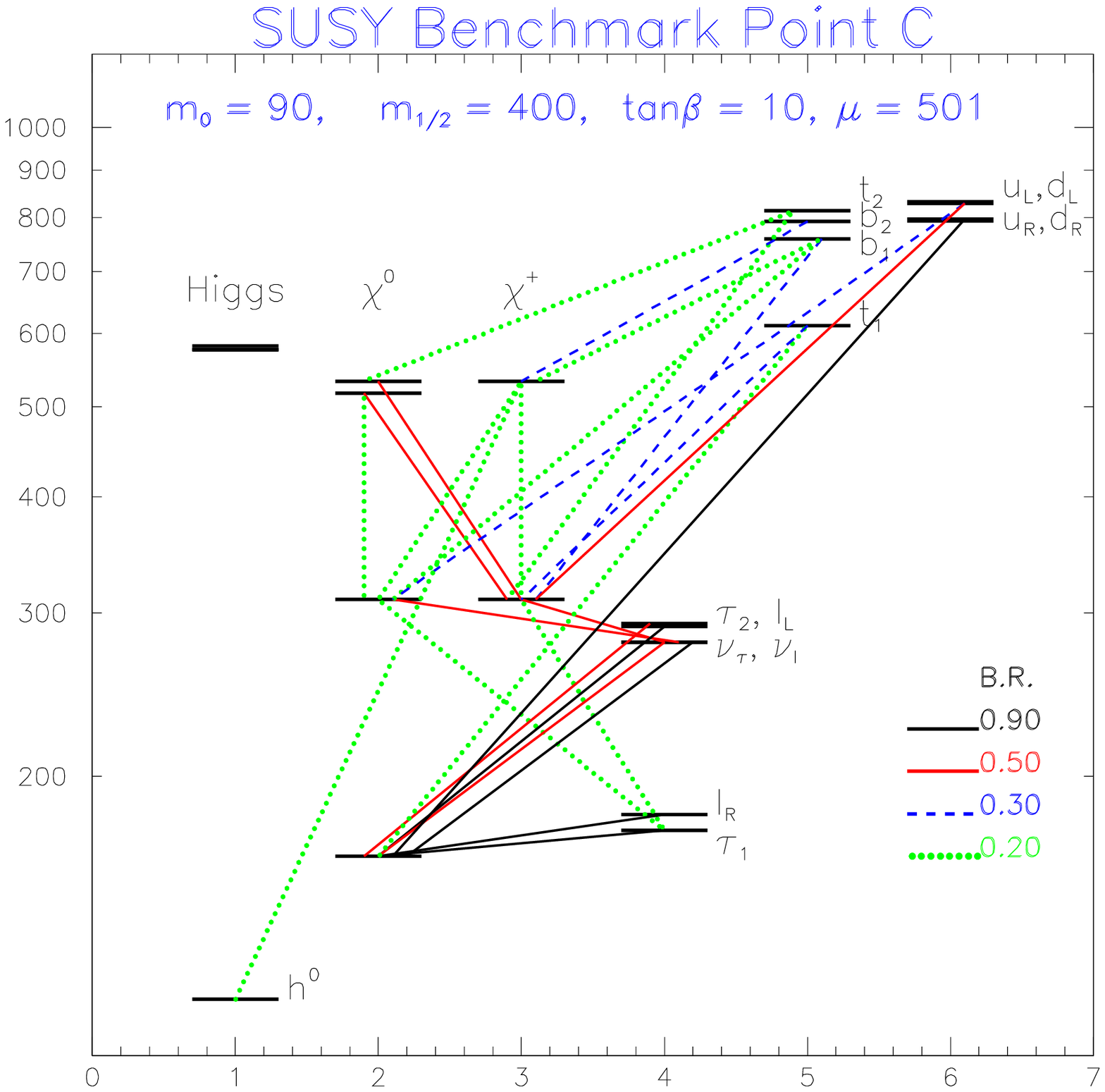}}
\resizebox{8cm}{!}{\includegraphics*[0,0][545,800]{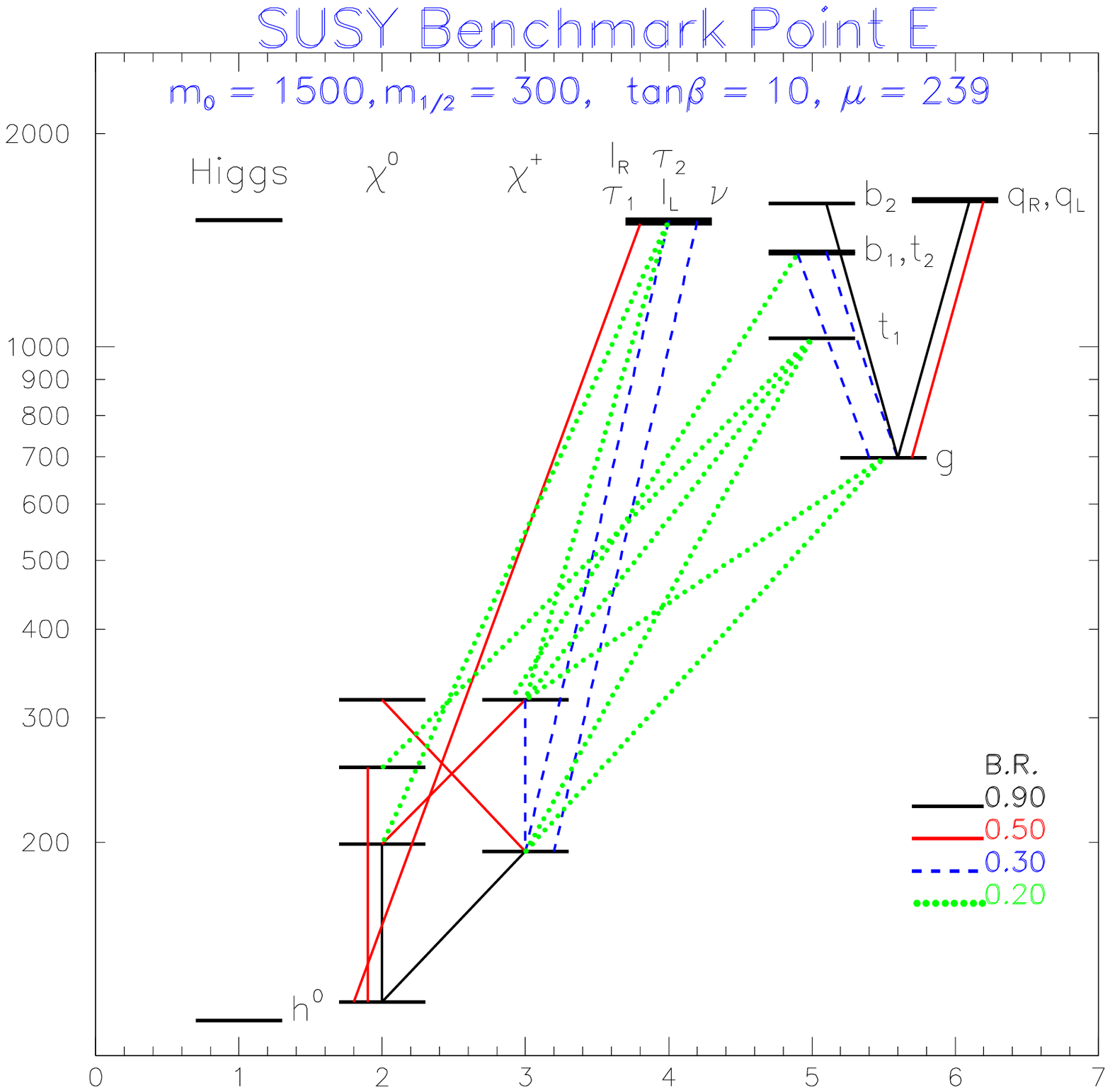}}
\vskip -1.5in
\resizebox{8cm}{!}{\includegraphics*[0,0][545,800]{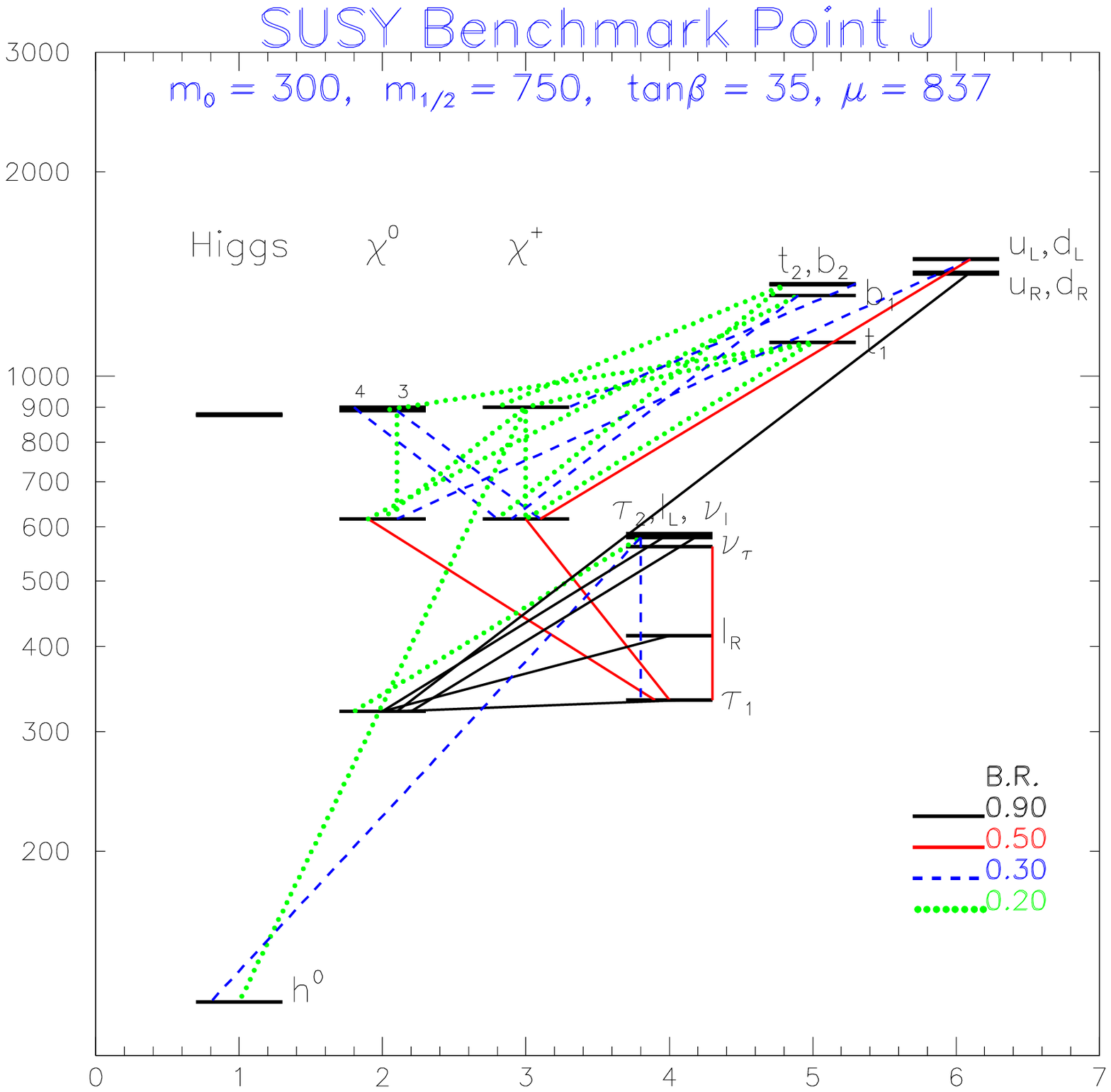}}
\resizebox{8cm}{!}{\includegraphics*[0,0][545,800]{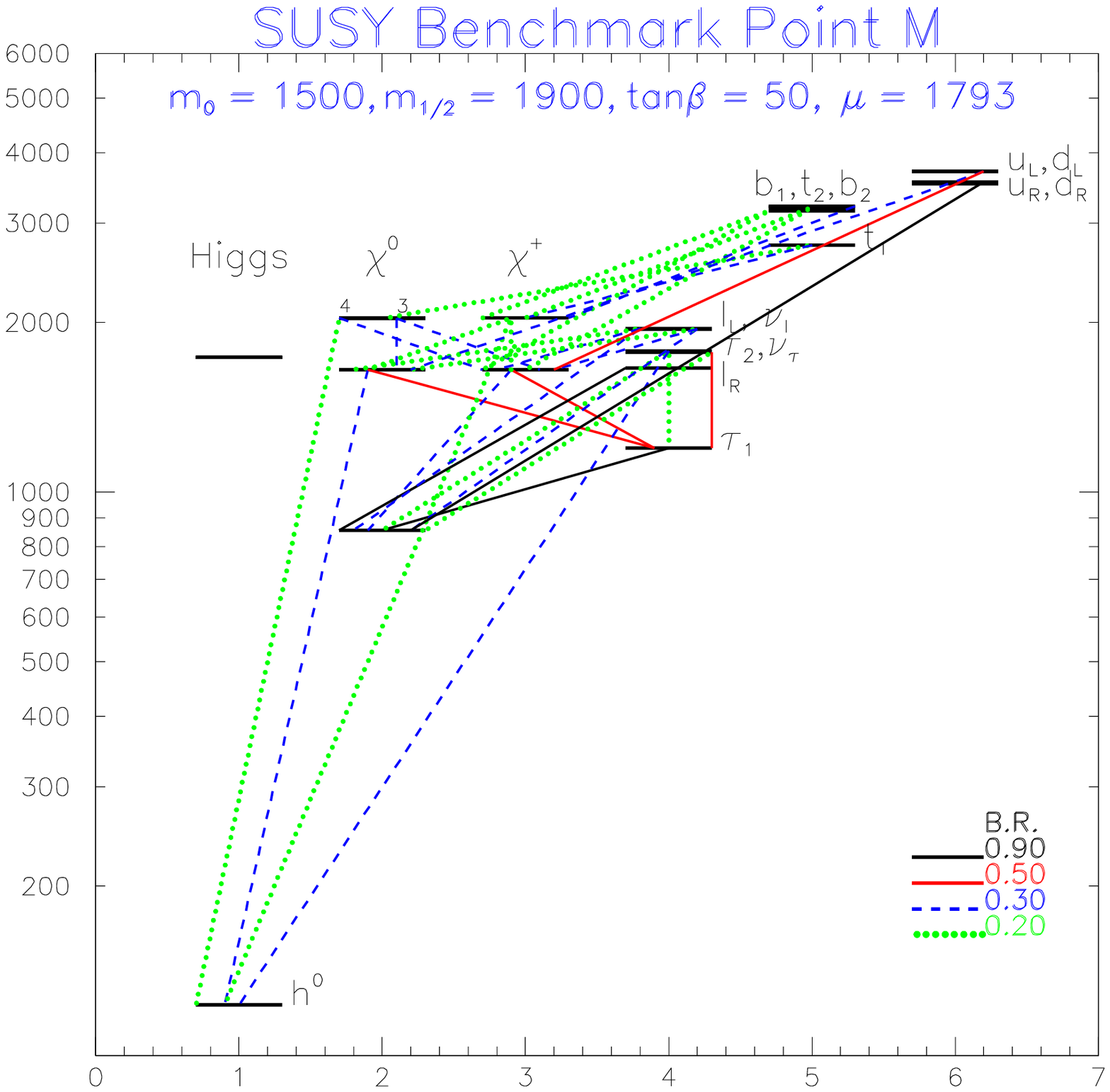}}
\vskip -0.5in
\caption{\label{fig:spectra}   
\it The supersymmetric spectra and principal decay
modes for benchmark points C,E,J and M, calculated using {\tt ISASUGRA
7.51}.}
\end{center}
\end{figure}

\begin{figure}[htb]
\begin{center}
\resizebox{8cm}{!}{\includegraphics*[0,0][545,800]{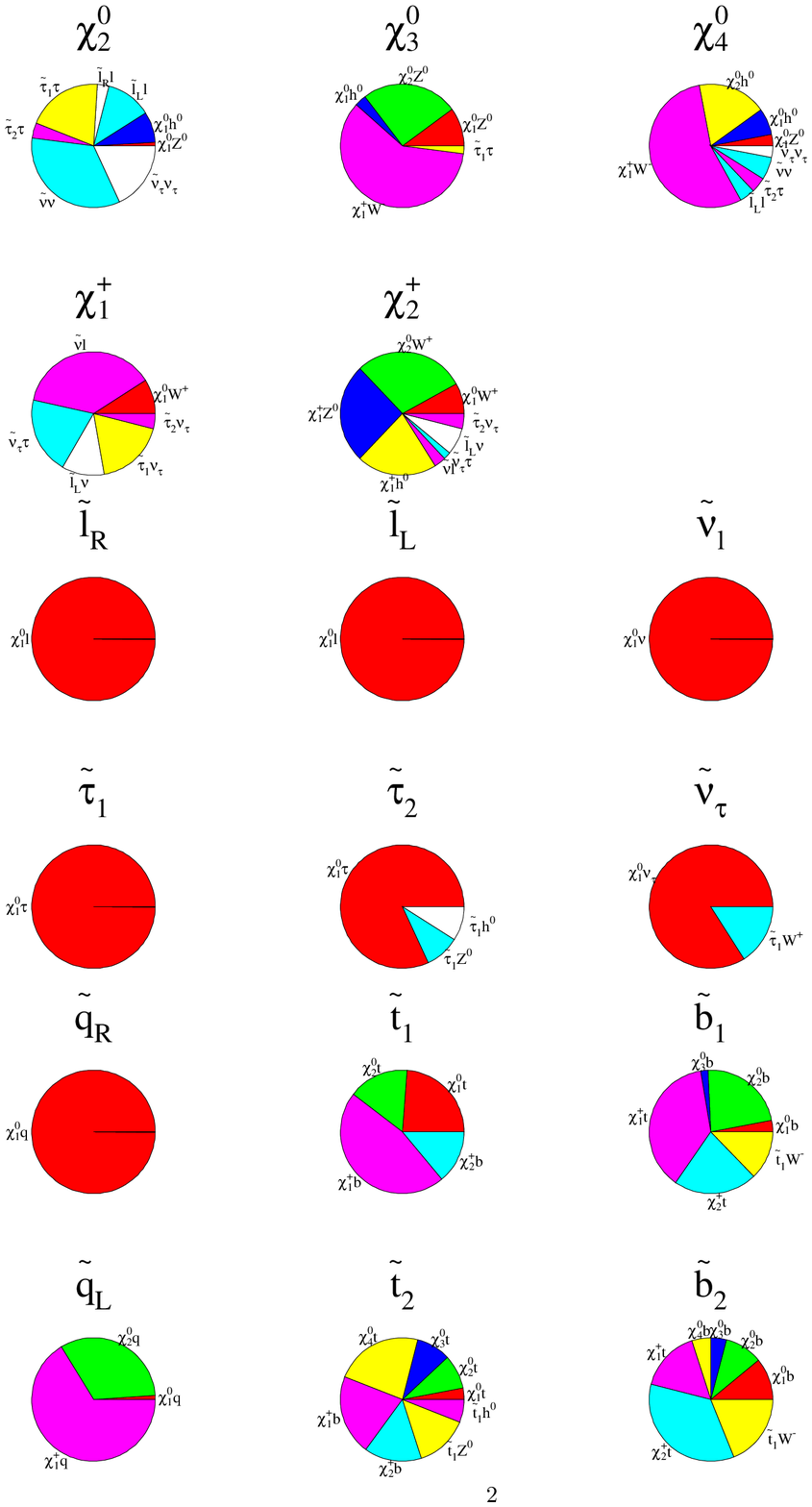}}  
\resizebox{8cm}{!}{\includegraphics*[0,0][545,800]{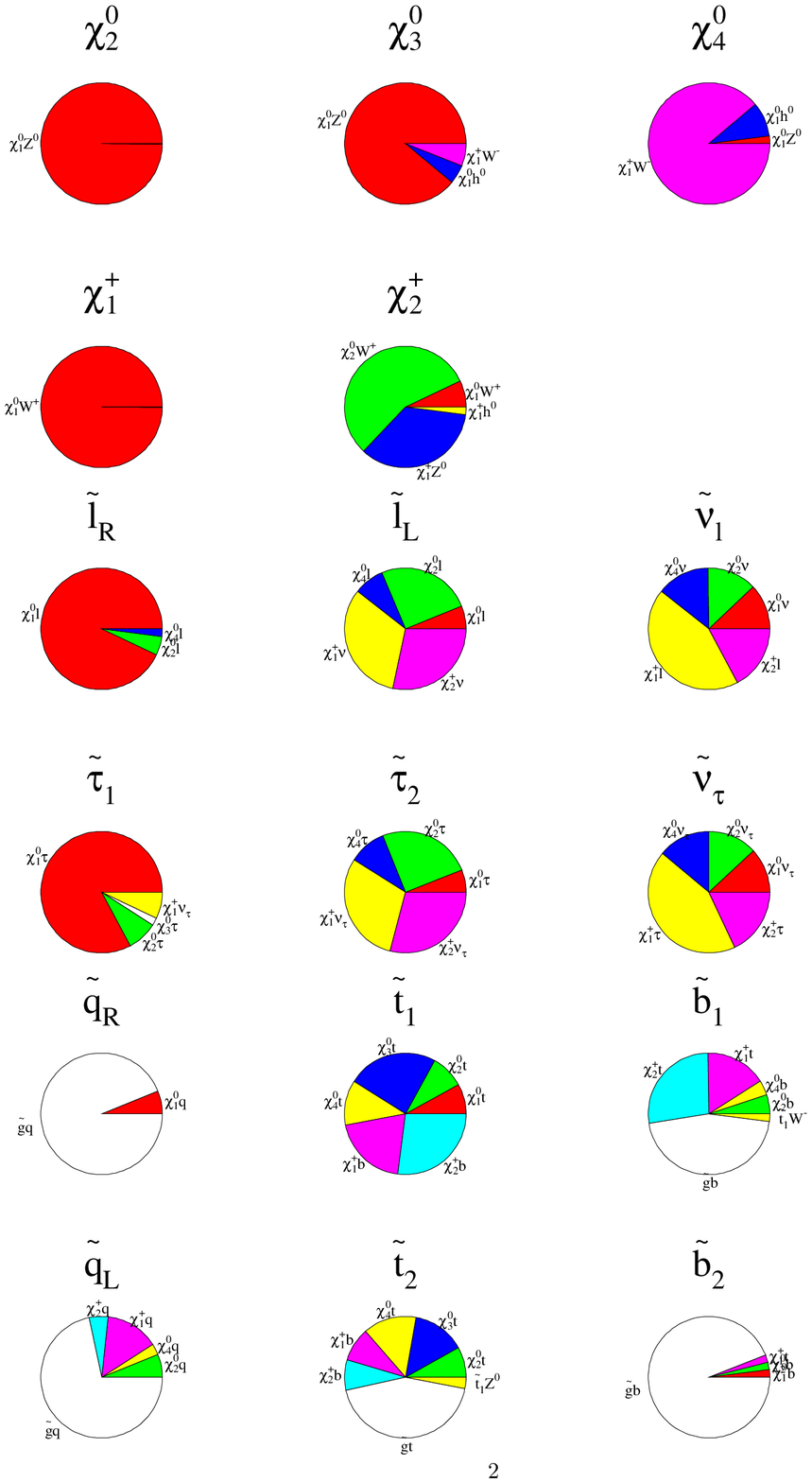}}  
\caption{
\label{fig:decays1}
\it Details of the principal decay branching ratios
for sparticles in benchmark points C and E, calculated using {\tt
ISASUGRA 
7.51}.}
\end{center}
\end{figure}

\begin{figure}[htb]
\begin{center}
\resizebox{8cm}{!}{\includegraphics*[0,0][545,800]{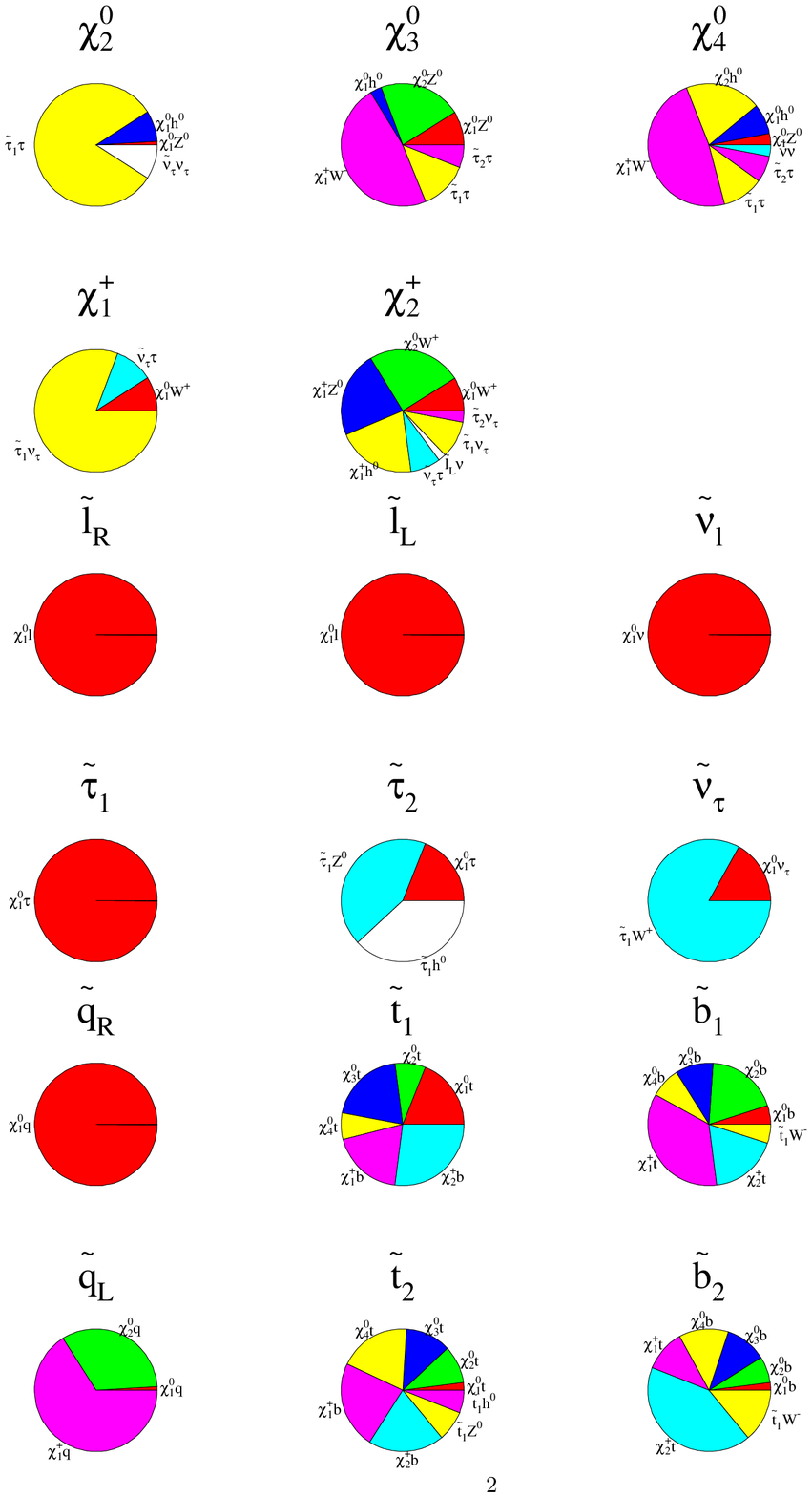}}  
\resizebox{8cm}{!}{\includegraphics*[0,0][545,800]{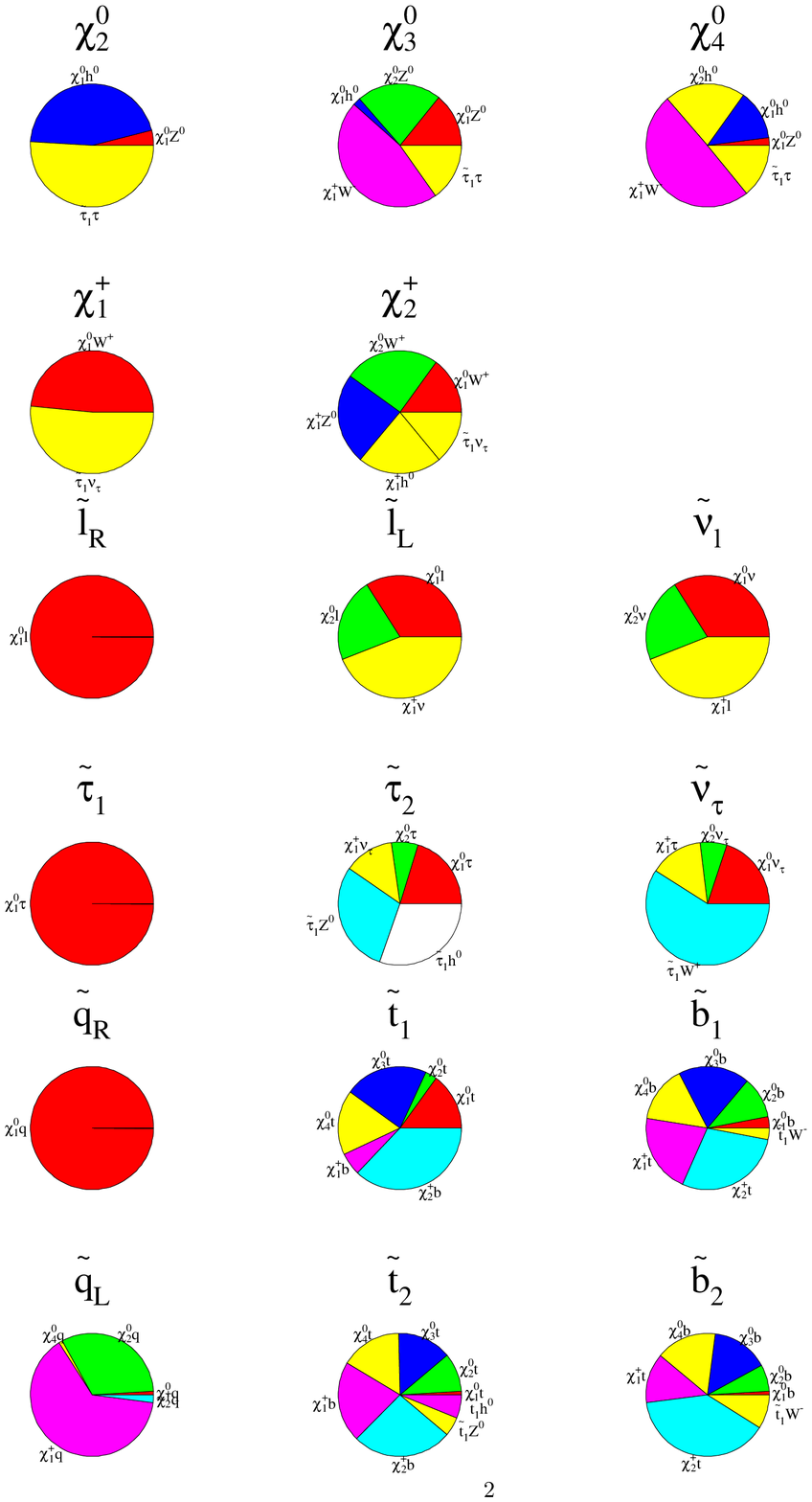}}  
\caption{\label{fig:decays2}
\it Details of the principal decay branching ratios
for sparticles in benchmark points J and M, calculated using {\tt
ISASUGRA 
7.51}.}
\end{center}
\end{figure}

Some examples of supersymmetric spectra in specific models are shown in
Fig.~\ref{fig:spectra}. Point C is in the `bulk' of the cosmological
region, in this case for $\tan \beta = 10$ and $\mu > 0$, point E is in
the focus-point region at large $m_0$, also for $\tan \beta = 10$ and $\mu
> 0$, point J is in the coannihilation `tail' for $\tan \beta = 35$ and
$\mu > 0$, and point M is in the rapid-annihilation `funnel' for $\tan
\beta = 50$ and $\mu > 0$.  Overviews of the dominant decay branching
ratios of the various sparticles in these models are shown in
Figs.~\ref{fig:decays1} and \ref{fig:decays2}.

\section{Prospective Supersymmetric Physics with Different Accelerators}

\begin{figure}
\epsfig{file=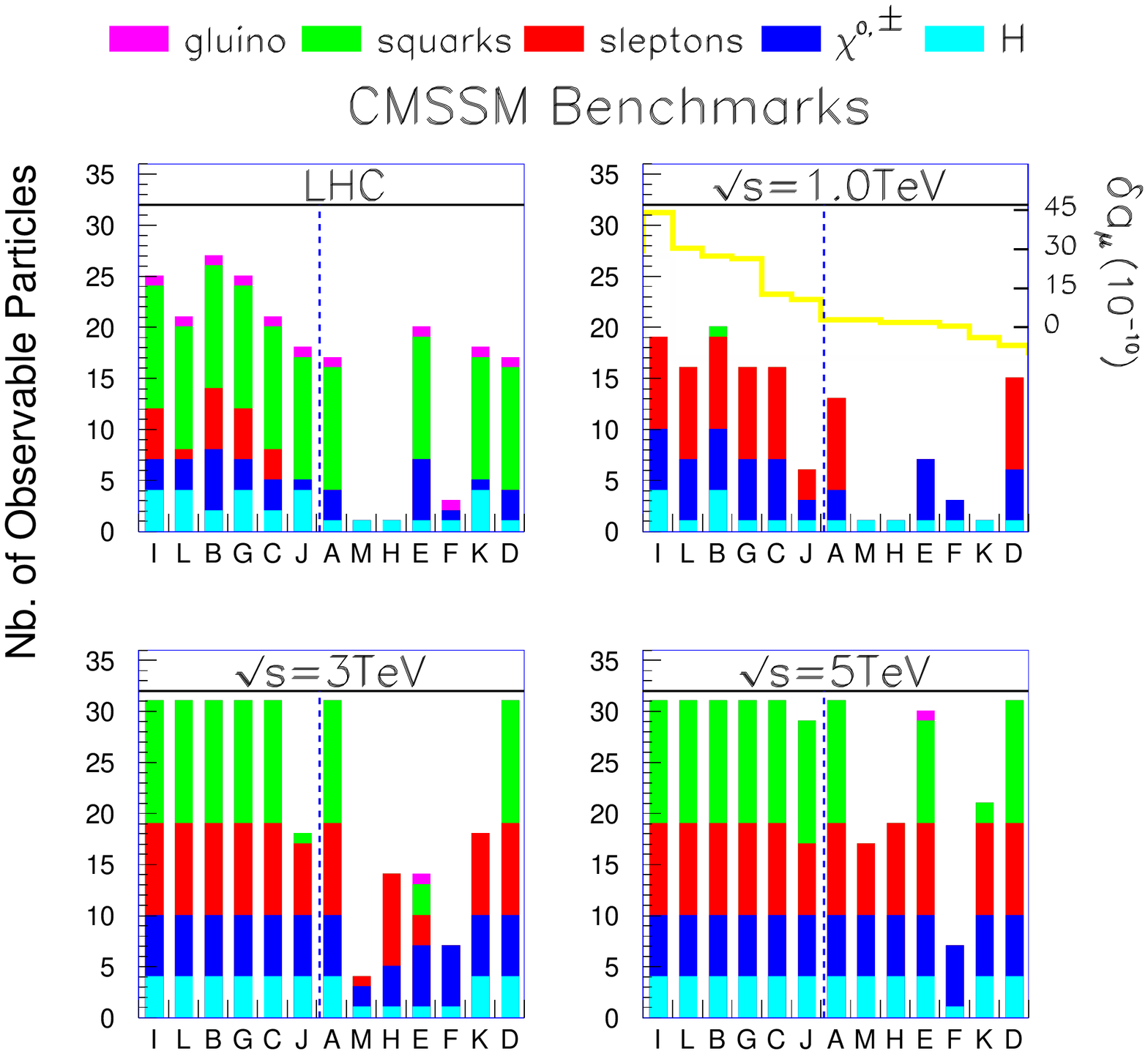,height=7in}
\vspace*{-0.5in}
\caption{\label{fig:Manhattan}
{\it Summary of the prospective sensitivities of the various accelerators
considered
to CMSSM production in the proposed benchmark scenarios, which are
ordered by their distance from the central value of $g_\mu - 2$, as
indicated by the pale (yellow) line in the second panel. We see clearly
the complementarity
between an $e^+ e^-$ collider and the LHC in the TeV range of energies,
with the former
excelling for non-strongly-interacting particles, and the LHC for
strongly-interacting sparticles and their cascade decays. CLIC provides
unparallelled physics reach for non-strongly-interacting sparticles,
extending beyond the TeV scale. We
recall that mass and coupling measurements at $e^+ e^-$ colliders
are usually much cleaner and more precise than at
hadron-hadron colliders such as the LHC. Note, in particular, that it is
not known how to distinguish the light squark flavours at the LHC.
}} 
\end{figure}

Estimates of the numbers of CMSSM particles accessible to different
accelerators in the various proposed benchmark scenarios are summarized in
Fig.~\ref{fig:Manhattan}. Caution should be used in the interpretation of
this figure, because it does not capture the different qualities of the
measurements at different colliders. For example, the masses and decay
modes of weakly-interacting sparticles and Higgs bosons can be measured
more precisely at $e^+ e^-$ colliders than at the LHC, if they are
kinematically accessible. Recall also that we have not chosen the points
such that they give a fair representation of the more and less likely
regions in the parameter space, but rather have opted to span the points
over as large and diverse a region as possible, to allow studies of the
consequences at the different colliders. Moreover, many of the
sensitivities assumed require verification and refinement by future
detailed studies. Hence Fig.~\ref{fig:Manhattan} as such cannot be used to
estimate if a particular collider would do well or not in discovering and
measuring supersymmetry, but rather illustrates the diversity of a
fraction of the possible scenarios and the complementarity of the
different colliders. 

The proposed benchmark points are ordered in Fig.~\ref{fig:Manhattan}
according to their degrees of compatibility with the recent measurement of
$g_\mu - 2$. Thus, if this measurement is confirmed as evidence for
physics beyond the Standard Model, the reader can see immediately which
benchmark scenarios are preferred, and how the prospects evolve for the
different accelerators considered.  One may also disfavour some of the
benchmark points because of the amounts of hierarchical and/or
cosmological fine-tuning they require, which we have documented in
Table~\ref{tab:derived_quantities}.

\subsection{Tevatron}

The Fermilab Tevatron has just begun its next run, which is planned to
deliver 1~fb$^{-1}$ of data per experiment per year in its first two
years, followed by a short shutdown for detector maintenance and
luminosity upgrade.  In the subsequent years, the Tevatron experiments are
hoping to collect as much as 5 ${\rm fb}^{-1}$ of data per experiment per
year, which might be enough for a Higgs (or supersymmetry) discovery at
the dawn of the LHC. We now assess the prospects for these searches in the
context of our proposed benchmark scenarios. 

\begin{itemize}

\item The search for the Higgs boson of the Standard Model is the
cornerstone of the Tevatron Run II program. A considerable amount of
effort has been put into optimising the Higgs discovery channels.  The
Higgs discovery reach in Run II is summarized in~\cite{Carena:2000yx}.
Based on the current expectations for the performance of the new
detectors, 3-$\sigma$ evidence for (5-$\sigma$ discovery of) a Standard
Model Higgs boson with a mass of 125 GeV is possible with somewhat less
that 10 (30) ${\rm fb}^{-1}$.  Similar conclusions apply to the lightest
CP-even Higgs boson of the MSSM, in the decoupling limit $m_A>>m_Z$, and
also for the CMSSM. Therefore, according to the Tevatron
study~\cite{Carena:2000yx}, this machine will be able to discover the
light Higgs boson $h^0$ in all of the benchmark points. 

\item In certain models with very light superpartners, the Tevatron also
has a shot at finding supersymmetry~\cite{Abel:2000vs}.  The gold-plated
mode for
supersymmetry discovery at the Tevatron is the clean trilepton channel
$3\ell\not\!\!E_T$.  The corresponding reach has been recently
re-evaluated in
\cite{Matchev:1999nb,Baer2}, with
improved background estimates and optimized analysis cuts.  In the case
where the two-body decays of $\tilde\chi^+_1$ and $\tilde\chi^0_2$ to
first generation sleptons are open, the 3-$\sigma$ reach extends up to
about
$m_{1/2}\sim 250$ GeV. Point B therefore appears to be on the edge of the
Tevatron sensitivity in this channel. However, by combining several
additional channels, e.g., the like-sign di-lepton
channel~\cite{Nachtman:1999ua,Matchev:1999nb}, the di-lepton plus tau jet
channel \cite{Baer:1998bj,Lykken:2000kp}, or the channels with jets,
$\not\!\!E_T$ and isolated leptons~\cite{Baer:1996ms}, as well as data
from both collaborations, point B might be observable with the full data
set.  In all the other 12 cases, however, the superpartners are too heavy
to be abundantly produced at the Tevatron, and will escape detection.
We should note that to date there have been no dedicated studies of the Tevatron
chargino/neutralino reach in the focus-point region, where {\em both}
chargino states are often kinematically accessible. 

\item A very interesting case is illustrated by point H.  The two lightest
supersymmetric particles, $\tilde\tau_1$ and $\tilde\chi^0_1$, are
extremely degenerate~\footnote{We recall that the high level of degeneracy
overcomes the kinematic suppression of the annihilation cross section due
to the relatively heavy LSP mass.}, and the $\tilde\tau_1-\tilde\chi^0_1$
mass gap is smaller than the tau mass $m_\tau$. This possibility can occur
anywhere along the borderline of the (red) shaded regions in
Figs.~\ref{fig:locations1} and \ref{fig:locations2}, where
$m_{\tilde\tau_1} -
m_\chi \rightarrow 0^+$. In such a case, the two-body decay
$\tilde\tau_1\rightarrow\tau\tilde\chi^0_1$ is closed, and, in the absence
of lepton-flavor violation in the slepton sector, the light tau slepton
predominantly decays via the four-body process
$\tilde\tau_1\rightarrow\ell\nu_\ell\nu_\tau\tilde\chi^0_1$, and is stable
on the scale of the size of the detector.  In that case, by looking for
long-lived massive charged particles \cite{Feng:1998zr,Culbertson:2000am},
one can probe slepton masses up to 225 GeV. The specific example of point
H falls beyond this projected sensitivity, but there exist points in the
CMSSM parameter space where this signature is accessible to the Tevatron. 

\item The lighter stop, ${\tilde t}_1$, is not within reach of the
Tevatron collider for any of our proposed scenarios. However, we recall
that we have always chosen $A_0 =0$, and that $m_{{\tilde t}_1}$ may be
reduced significantly for some other values of $A_0$
\cite{Demina:2000ty}. 

\end{itemize}

This brief discussion shows that the Tevatron collider has good prospects
for discovering the lightest CMSSM Higgs boson within all our proposed
benchmark scenarios, and some prospects for detecting supersymmetric
particles. In the latter searches, our benchmark points suggest some
unconventional scenarios that should be kept in mind. 

\subsection{LHC}

A preliminary inspection has been made of the LHC potential for these
benchmark points, based on the simulation results summarized in the ATLAS
Physics Technical Design Report \cite{ATLASTDR} and in the CMS Note
\cite{CMS98006}~\footnote{Previous particle-level estimates of the LHC
sensitivity can be traced from~\cite{BCDPT}.}. A detailed study is clearly
required before a real
assessment of the LHC physics potential for these benchmarks can be made.
For a preliminary look, the following assumptions were adopted to estimate
the discovery potential of the LHC, assuming ATLAS+CMS combined, together
with an integrated luminosity of 300 $fb^{-1}$ per experiment. 

\begin{itemize}

\item The light Higgs is always within reach, 
since its production rate
and decay signatures are very similar to those in the Standard Model,
as demonstrated by detailed studies of Standard Model and MSSM
light Higgs bosons at LHC.

\item The heavy neutral Higgses can be searched for via their decays
$H^0,A^0 \rightarrow \tau \tau$ and $\mu \mu$, and the quoted $H^{\pm}$
discovery range is based on the decay $H^{\pm} \rightarrow \tau \nu$ and
${\bar b} t$. As is well known, there is a region at large $m_A$ and
moderate $\tan \beta$ where the heavier Higgs bosons may escape detection
at the LHC, and this `hole' is reflected in our
results~\footnote{Supersymmetric decays of the heavier Higgs bosons might
also be an interesting signature.}. 

\item The observation of gauginos is either through direct Drell-Yan
production of $\chi_2^0 \chi_1^{\pm}$, leading to trilepton final states,
or via the inclusive production of charginos and neutralinos in the decays
of squarks and gluinos.  Neutralino decays to same-flavour di-leptons and
missing energy yield a characteristic end point at the upper edge of the
mass spectrum.  If the leptonic decays of gauginos are enhanced because the
sleptons are light, according to the published study \cite{CMS98006}, the
$5-\sigma$ discovery
region for $\chiz_2$ is bounded approximately by $m_{1/2} \leq 200$ GeV or
$m_0 \leq 0.45 m_{1/2}$ for $m_{1/2} \leq 900$ GeV.
However, it should be noticed that, if the squark and gluino masses are
light
enough to yield sufficient production rates, charginos and neutralinos
with $W$- and $Z$-like branching ratios are also observable from cascade
decays (which applies to points E and L). The treatment of the
major background to the clean trilepton channel in those studies may now be
updated along the lines of~\cite{Matchev:1999nb,Baer2}. Heavier gaugino
states are assumed observable in the same region of $(m_{1/2}, m_0)$,
provided their mass is less than 450 GeV.  More detailed studies are needed
to assess the observability of the various gauginos for the various
benchmark points, since the results depend strongly on the involved masses
and decay patterns.  

\item The squarks have conservatively been considered
observable when $m_{\tilde{q}} \leq 2500$ GeV and $m_{\tilde{t}_1} \leq
2000$ GeV.  However, we note that it is not known how to distinguish the
different light squark flavours at the LHC, and that the separation of
squarks from gluinos needs to be studied in more detail. 

\item The
charged sleptons can be observed through their direct production up to
masses of $\sim 350$~GeV and perhaps indirectly in the decays of other
supersymmetric particles up to masses of $\sim 250$~GeV. In addition, we
have assumed that they are observable only if the mass differences
$m_{\tilde \ell} - m_{\chiz_1} \geq 30$~GeV.

\item As was mentioned earlier, the ${\tilde \tau}_1$ may be
long-lived, if $m_{1/2}$ and $m_0$ have values close to the boundary where
the ${\tilde \tau}_1$ becomes the LSP, as for point H. The LHC
collaborations have already made analyses of their sensitivities for such
scenarios, motivated by gauge-mediated models. They have found that such a
${\tilde \tau}_1$ can be detected at the LHC by measuring delayed signals
produced by these particles in the external muon spectrometers of
ATLAS~\cite{ATLPHYSCOM2000/013} and CMS~\cite{CMSCR1999/019}, whose time
resolutions are $\sim 1$~ns.  By combining time and momentum measurements,
it is possible to determine $m_{{\tilde \tau}_1}$. A mass resolution of
$\sim 3.5$\% was obtained for gauge-mediated point G2b in~\cite{ATLASTDR},
where $m_{{\tilde \tau}_1} \sim 100$~GeV. At point H, the ${\tilde \tau}_1$
is much heavier, and its detectability requires further study. 

\item The
sneutrinos are not counted as observable, since they have large branching
ratios for invisible decays into $\tilde{\nu} \rightarrow \nu \chi_1^0$. 

\end{itemize}

Our preliminary estimates of the numbers of detectable particles of each
kind for each benchmark point are summarized in
Table~\ref{tab:msugra_obsLHC}.

\begin{table}[htb]
\centering
{\bf Prospective observability at the LHC}\\
{~}\\
\begin{tabular}{|c||r|r|r|r|r|r|r|r|r|r|r|r|r|}
\hline
    Model      &  A   &  B   &  C  &  D   &  E    &  F   &  G  &  H   &  I  &  J  &  K   &  L  &  M      \\ 
\hline
    $m_{1/2}$& 600  & 250 & 400 & 525  &  300& 1000& 375 & 1500& 350  & 750  & 1150 & 450& 1900   \\
    $m_0$    & 140  & 100 &  90 & 125  & 1500& 3450& 120 & 419 & 180  & 300  & 1000 & 350& 1500  \\
    $\tan{\beta}$
              & 5    & 10  & 10  & 10   & 10  & 10  & 20  & 20  & 35   & 35   & 35   & 50 & 50     \\
    sign($\mu$)& $+$  & $+$ & $+$ & $-$  & $+$ & $+$ & $+$ & $+$ & $+$  & $+$  & $-$  & $+$& $+$    \\ 
\hline
    $h^0,H^0,A$& 1    & 1    & 1   & 1    & 1     & 1    & 3   & 1    & 3   & 3   & 3    & 3   & 1      \\
    $H^{\pm}$  & 0    & 1    & 1   & 0    & 0     & 0    & 1   & 0    & 1   & 1   & 1    & 1   & 0     \\
    $\chi^0_i$/$\chi^{\pm}_j$
               & 3    & 6    & 3   & 3    & 6     & 1    & 3   & 0    & 3   & 1   & 1    & 3   & 0     \\
    sleptons   & 0    & 6    & 3   & 0    & 0     & 0    & 5   & 0    & 5
& 0   & 0    & 1   & 0     \\
    squarks    & 12   & 12   & 12  & 12   & 12    & 0    & 12  & 0    & 12  & 12  & 12   & 12  & 0     \\
    gluino     & 1    & 1    & 1   & 1    & 1     & 1    & 1   & 0    & 1   & 1   & 1    & 1   & 0     \\
\hline
\end{tabular}
\caption[]{\it Numbers of particles for each benchmark model thought to
be accessible at
the LHC. The observabilities we assume are obtained by extrapolating from
previous simulation studies by ATLAS and CMS.
}
\label{tab:msugra_obsLHC}
\end{table}

The great potential of the LHC for the discovery of squarks and the gluino
as well as the lightest Higgs is clearly apparent.  For most benchmark
points, the discovery of some of the gauginos and sleptons would also be
possible. However, it is also seen that for the points with the heaviest
spectra, namely models F, H and M, the LHC may have difficulty in finding
any MSSM state except the light Higgs boson.  It is, however, not excluded
that at least some of the states in these solutions will be found to be
detectable after more detailed work has been done, as some squark masses
are close to the limits chosen above. There is also the possibility of a
luminosity upgrade for the LHC, which might extend its reach for squark
and gluino masses up to about 3 TeV. We have not studied the extent to
which the different squark flavours may be distinguished at the LHC. 

Previous studies\cite{ATLASTDR,CMS98006} demonstrated for six benchmark
points that several sparticle spectroscopic parameters could be
measured with precisions of a few percent using kinematic
distributions.  This allows the fundamental parameters of the CMSSM to be
constrained to $1\% - 10\%$, depending on the point studied. 

\subsection{Linear $e^+ e^-$ Colliders up to 1 TeV}

Electron-positron collisions at centre-of-mass energies up to about 1~TeV
are
envisaged by the TESLA~\cite{TESLATDR2}, NLC~\cite{NLC,Abe:2001mv} and
JLC~\cite{JLC} projects. In each
case, a possible
first phase at energies up to about 500~GeV is also considered. In both
the lower- and higher-energy phases, data samples of the order of
1000~fb$^{-1}$ or more will be collected over a period of several years.
Recent supersymmetric benchmark studies for TESLA~\cite{TESLATDR} have
included two CMSSM scenarios with $\tan {\beta}$ = 3 (30), $m_0 = 100
(160)$~GeV, $m_{1/2}=200$~GeV, $\mu>0$, and $A_0$= 0 (600) GeV. In the
case of the first point, the Higgs mass is less than 100 GeV, and thus
now ruled out by LEP. 

Sparticles can be produced at any linear $e^+ e^-$ collider if its
centre-of-mass energy is larger than twice the mass of the sparticles, the
pair production threshold, except for heavier charginos and neutralinos,
which can be produced in association with the lightest chargino or
neutralinos, respectively~\footnote{For sufficiently light neutralinos and
sneutrinos, observation of the radiative production of otherwise invisible
final states may be experimentally accessible.}. Typical supersymmetric
signals are multi-lepton final states and multi-jet final states with
large missing transverse energy.  Sneutrinos can be detected at threshold
energies if these can decay into channels including charged leptons with a
sufficiently large branching ratio.  For example, in some scenarios the
${\tilde \nu}_{\tau}$ can decay into $\tau {\tilde W}$.  Otherwise,
sneutrinos can be detected at higher energies via the decays of charginos.
Apart from the threshold requirement to produce the particles, we require
the branching ratio times cross section for the detectable channels to be
larger than 0.1 fb in order to observe the sparticle, leading to at least
100 produced sparticles in the total data sample. 

Table~\ref{tab:msugra_obsECOL} shows the sparticles which can be observed
at a 1 TeV linear $e^+ e^-$ collider in each of the different benchmark
points proposed in this paper. The unpolarized cross sections and the
decay branching ratios were computed using {\tt ISASUGRA 7.51}.  The
listed numbers of observable particles take into account the decays of
sneutrinos and neutralinos into undetectable final states.

\begin{table}[htb]
\centering
{\bf Observable particles at linear $e^+ e^-$ colliders}\\
{~}\\
\begin{tabular}{|l|c||r|r|r|r|r|r|r|r|r|r|r|r|r|}
\hline
$\sqrt{s}$
    &Model    &  A   &  B  &  C  &  D   &  E  &  F  &  G  &  H  &  I   &  J   &  K   &  L &  M    \\ 
\hline
    &$m_{1/2}$& 600  & 250 & 400 & 525  &  300& 1000& 375 & 1500& 350  & 750  & 1150 & 450& 1900   \\
    &$m_0$    & 140  & 100 &  90 & 125  & 1500& 3450& 120 & 419 & 180  & 300  & 1000 & 350& 1500  \\
    &$\tan{\beta}$
              & 5    & 10  & 10  & 10   & 10  & 10  & 20  & 20  & 35   & 35   & 35   & 50 & 50     \\
    &sign($\mu$)& $+$  & $+$ & $+$ & $-$  & $+$ & $+$ & $+$ & $+$ & $+$  & $+$  & $-$  & $+$& $+$    \\ 
\hline

1.0 & Higgs   & 1    & 4   & 1   & 1    & 1   & 1   & 1   & 1   & 4    & 1    & 1    & 1  & 1      \\
1.0 & $\chi^{0,\pm}_i$
              & 3    & 6   & 6   & 5    & 6   & 2   & 6   & 0   & 6    & 2 
& 0    & 6  & 0      \\
1.0 & slept   & 9    & 9   & 9   & 9    & 0   & 0   & 9   & 0   & 9    & 3    & 0    & 9  & 0      \\
1.0 & squa    & 0    & 1   & 0   & 0    & 0   & 0   & 0   & 0   & 0    & 0    & 0    & 0  & 0      \\ 
\hline 

3.0 & Higgs   & 4    & 4   & 4   & 4    & 1   & 1   & 4   & 1   & 4    & 4    & 4    & 4  & 1      \\
3.0 & $\chi^{0,\pm}_i$
              & 6    & 6   & 6   & 6    & 6   & 6   & 6   & 4   & 6    & 6
& 6    & 6  & 2      \\
3.0 & slept   & 9    & 9   & 9   & 9    & 3   & 0   & 9   & 9   & 9    & 7
& 8    & 9  & 1      \\
3.0 & squa    & 12   & 12  & 12  & 12   & 3   & 0   & 12  & 0   & 12   & 1 
& 0    & 12 & 0      \\ 
\hline

5.0 & Higgs   & 4    & 4   & 4   & 4    & 4   & 1   & 4   & 4   & 4    & 4    & 4    & 4  & 4      \\
5.0 & $\chi^{0,\pm}_i$
              & 6    & 6   & 6   & 6    & 6   & 6   & 6   & 6   & 6    & 6    & 6    & 6  & 6      \\
5.0 & slept   & 9    & 7   & 9   & 9    & 9   & 0   & 9   & 9   & 9    & 7    & 9    & 9  & 7      \\
5.0 & squa    & 12   & 12  & 12  & 12   & 10  & 0   & 12  & 0   & 12   &
12   & 2    & 12 & 0      \\
\hline
1.0 & TOT     & 13   & 20  & 16  & 15   & 7   & 3   & 16  & 1   & 19   & 6 
& 1    & 16 & 1      \\
3.0 & TOT     & 31   & 31  & 31  & 31   & 13  & 7   & 31  & 14  & 31   &
18 & 18   & 31 & 4     \\
5.0 & TOT     & 31   & 29  & 31  & 31   & 29  & 8   & 31  & 19  & 31   &
29 & 21   & 31 & 17     \\
\hline
\end{tabular}
\caption[]{\it
Numbers of particles accessible for each benchmark model for various
lepton-antilepton collider centre-of-mass energies in TeV.
Channels are considered observable when their cross section
times branching ratio to visible final states exceeds 0.1 fb,
taking account of the invisible final states originating from some
neutralino and sneutrino decay modes.
No considerations of realistic detection efficiencies have been included.
}
\label{tab:msugra_obsECOL}
\end{table}

We note the following points concerning searches for CMSSM particles at a 
linear $e^+ e^-$ collider in the energy range $\lappeq 1$~TeV.

\begin{itemize}

\item The lightest CMSSM Higgs boson is always detectable in a first phase
even below $\sqrt{s} \sim 500$~GeV, and its mass, width, spin-parity and
couplings can be measured with high precision~\cite{TESLATDR}. In the
second phase, interesting measurements could be made of the trilinear
Higgs self-coupling. These possibilities would go far beyond the Higgs
measurements possible at the Tevatron or the LHC, and could be used to
constrain significantly the CMSSM parameter space. Higgs studies would
form an attractive cornerstone of the first phase of a linear $e^+ e^-$
collider.

\item
In all the proposed scenarios except H, K and M, a number of additional
sparticles is within reach of a 1~TeV linear $e^+ e^-$ collider. These
exceptions are amongst those disfavoured by the measurement of $g_\mu - 2$. 
In approximately half of the cases, particularly those which are
consistent with the $g_\mu - 2$ measurement at the 2 $\sigma$ level, a
large fraction of the sleptons and gauginos can be detected. In
particular, points B and I are very favourable for a linear collider with
centre-of-mass energy up to 1 TeV.

\item
The LHC and such a linear collider are largely complementary in their
sparticle mass reaches.  For example, in the case of point L, the linear
collider would see all the sleptons, while the LHC would see all the
squarks. 

\item
In contrast to the LHC, one notices that the squarks are not
generally accessible in the proposed models.

\item
Beyond the discovery of sparticles, a crucial issue in the understanding
of the nature of any new physics observed will be the accuracy obtainable
in the determination of the sparticle masses and decays, and also their
quantum numbers and mixing.  A strong advantage of lepton colliders is the
{\it precision} with which such sparticle properties can be measured. 
Typically, the masses of sleptons and gauginos can be determined with a
precision of a few per mille, by threshold
scans and by measuring end points of two-body decay channel signatures in
inclusive distributions. This uncanny precision, even for a limited number
of sparticles, will be of cardinal importance for the reconstruction of
the underlying supersymmetric model and breaking mechanism~\cite{Blair}. 

\item
Furthermore, the availability of polarized beams at a linear collider will
provide additional tools for identifying supersymmetric particles and
allow for additional measurements of parameters of the supersymmetric
model, such as the mixing angles of the sparticles~\cite{TESLAPOL}. 

\item
In a few cases the heavy Higgs particles are within reach of a 1 TeV
$e^+e^-$ collider, but we also note that the reach for heavy Higgses can
be extended with a photon collider option~\cite{Boos}, in which the lepton
beams are
converted into high-energy photon beams, with a peak energy of up to 80\%
of the original incoming lepton beam energy.  Heavy Higgses can be singly
produced in two photon interactions, increasing the reach in detection to
Higgs particles with masses of approximately 75-80\% of the centre-of-mass
energy. Such an option would be very useful for the points C, D, G and L
where the heavy Higgs particles become accessible in the two-photon
collider mode.

\item
We note also that a precise measurement of the two-photon
width of the light Higgs boson, which appears feasible at a photon
collider~\cite{Soldner}, might give 
evidence of the existence of new physics, such as supersymmetry, even
if no other new particles besides the Higgs boson have been discovered at 
the LHC or a 1 TeV linear collider.

\end{itemize}

\subsection{CLIC}

CLIC is a project for an $e^+e^-$ linear collider which aims at a
centre-of-mass energy of 3 TeV, upgradable to 5 TeV at a later stage.  The
luminosity will be about 10$^{35}$cm$^{-2}$s$^{-1}$. To achieve this
luminosity, CLIC will work in the high beamstrahlung regime~\cite{CLIC}.
The physics potential of this machine is currently being
studied~\cite{CLICphys}. 

To obtain a first, very crude, estimate of the potential of a linear $e^+
e^-$ collider in the multi-TeV range, we have used the same criteria as in
the previous section, i.e., we assume that sparticles can be detected
provided their production cross section multiplied by their branching
ratio to visible final states exceeds 0.1 fb~\footnote{This limiting cross
section is why some sparticles are not counted as observable, even though
they are kinematically accessible. For example, at 5~TeV , at point K the
${\tilde d}_R$ and ${\tilde s}_R$ have cross sections of
only 0.01 fb and the ${\tilde b}_2$ only 0.04 fb, whilst at point H the
${\tilde t}_1$ is produced with a
cross section of only 0.03 fb.}. CLIC is estimated to yield an integrated
luminosity of about 1000 fb$^{-1}$ in a year, of which about a third would
be close to the nominal $E_{CM}$. Therefore, one could accumulate 1000
fb$^{-1}$ at this nominal energy in about three years, corresponding to
100 events produced for a cross section of 0.1 fb. More accurate estimates
of the CLIC sensitivity will become possible only after simulation of the
signals and backgrounds.  Preliminary studies indicate that, although CLIC
will operate with more beamstrahlung and hence a less well-defined
centre-of-mass energy than lower-energy linear colliders, there are
several physics topics for which the accelerator environment
is not a serious disadvantage~\cite{CLICphys}.

The unpolarized cross sections and the decay branching ratios were again
computed using {\tt ISASUGRA 7.51}.  The numbers of observable particles,
after taking into account the decays of sneutrinos and neutralinos into
undetectable final states, are also summarized in
Table~\ref{tab:msugra_obsECOL}.
In preparing this Table, the possibility of detecting sneutrinos from the
two-body decays of charginos has been taken into account, and we note
that, in scenario B, the sneutrinos are only observable through their
associated production. Note that 
due to the decrease of the cross section as
function of energy, the $\sNu_{\mu}$ and $\sNu_{\tau}$ which are
observable at 3 TeV but are in principle no longer at 
an energy at 5 TeV if the luminosity is the same
as for a 3 TeV collider. However this would ignore the fact that
the machine could run at lower energy, and also the fact that CLIC makes
an `autoscan', by virtue of the beamstrahlung photon spectrum.
Hence, for the counting in Table~\ref{tab:msugra_obsECOL},
these particles are considered to be detectable at a collider which
can go up to 5 TeV. 
Finally, we note that, at point E, the gluinos are observable at
energies $\geq$ 3 TeV, as they constitute the dominant decay mode of the
squarks, but they are not listed in the Table.

Assuming that the LHC and an $e^+ e^-$ linear collider in the range
$\lappeq 1$~TeV have been taking data for several years before the start
of a 3 TeV machine like CLIC (CLIC3000), we infer that supersymmetry will
most probably already have been discovered by that time, if it exists. 
Hence the r\^ole of CLIC may consist mainly of completing the sparticle
spectrum, and disentangling and measuring more precisely the properties of
sparticles already observed at the LHC and/or a lower-energy $e^+ e^-$
linear collider. However, a machine like CLIC would be needed even for the
direct discovery of supersymmetry in the most problematic cases, namely
scenarios H and M.

A few benchmark points emerge as typical of situations which could arise
in the future. 

\begin{itemize}

\item Point C has very low masses, and is representative also of points A,
B, D, G, I, L.  In these cases, LHC would have discovered the $H^{\pm}$,
as well as seen the $h^0$, and also the gauginos $\chiz_1$, $\chiz_2$ and
$\chipm_1$, some of the charged sleptons, the squarks and the gluino.  A
1~TeV
linear collider would enable the detailed study of the $h^0$ and of the
same gauginos and sleptons, and it might discover the missing
sleptons and gauginos in
some of the scenarios. However, one would require CLIC, perhaps running
around 2 TeV, to complete the particle spectrum by discovering and
studying the heavy Higgses and the missing gauginos.  CLIC could also
measure more precisely the squarks and in particular disentangle the left-
and right-handed states and, to some extent, the different light squark
flavours.

\item Point J features intermediate masses, rather similar to point K. 
Here, the LHC would have discovered all the Higgs bosons, the squarks and
the gluino, but no gauginos nor sleptons.  The 1~TeV $e^+ e^-$ linear
collider would study in detail the $h^0$ and could discover the $\sEl_R$,
$\sMu_R$ and $\sTau_1$, but other sparticles would remain beyond its
kinematic reach.  CLIC3000 could then study in detail the heavy Higgses. 
It would also discover and study the gauginos and the missing sleptons,
and even observe in more detail a few of the lighter squarks that had
already been discovered at LHC. However, to see the remaining squarks at a
linear collider would require CLIC to reach slightly more than 3 TeV. 

\item Point E has quite distinctive decay characteristics, due to the
existence of heavy sleptons and squarks.  In this situation, the LHC would
have discovered the $h^0$, all squarks and the gluino.  The gauginos are
in principle accessible, but their discovery may be made more difficult
because their predominant decays are into jets, contrary to the previous
benchmark points, and sleptons would remain unobserved.  At a 1~TeV $e^+
e^-$ linear collider, the detailed study of the $h^0$ and of the gauginos
could be undertaken.  The discovery of the first slepton, actually a
$\sNu_e$, could be made at CLIC3000, which could also study the three
lightest squarks.  The discovery and analysis of the heavy Higgses would
then require the CLIC energy to reach about 3.5 TeV, which would also
allow the discovery of all sleptons and the observation of all squarks.

\item Point M has quite heavy states, rather close to scenario H.  The LHC
would only discover the $h^0$, all other states being beyond its reach. 
The existence of supersymmetry might remain an open question!  The 1~TeV
$e^+ e^-$ linear collider also sees only the $h^0$, but can study it in
detail, and so might provide indirect hints of supersymmetry.  A CLIC3000
would be able to discover most of the gauginos and the ${\tilde \tau}_1$,
but CLIC with 3.5 to 4 TeV would be required to discover the remaining
gauginos and sleptons. To discover the squarks, $\ell^+ \ell^-$ collisions
above 6.5 TeV would be needed. There is currently no $e^+ e^-$ project
aiming at such energies, and neutrino radiation becomes a hazard for
$\mu^+ \mu^-$ colliders at such energies. 

\item Point F also has heavy states.  Here again, LHC sees only the $h^0$,
but would also find the gluino. The lightest gauginos would be within the
reach of a 1~TeV $e^+ e^-$ linear collider, and the heavier ones within
the reach of CLIC3000.  But it would only be with $\ell^+ \ell^-$
collisions in the 6.5-7.5 TeV region that the heavy Higgses, the sleptons
and the squarks could be found.

\item As in the 1~TeV $e^+ e^-$ linear collider case, a photon collider
option for CLIC would extend the discovery range for heavy Higgs bosons. 
It would additionally allow one to discover all four Higgs bosons in
scenarios E, H and M, for a 3 TeV collider, and also in F, for a 5 TeV
Collider. 

\item We also note that polarization would have advantages similar to
those discussed earlier for a lower-energy $e^+ e^-$ linear collider.
 
\end{itemize}

\section{Conclusions and Prospects}

We have proposed some benchmarks that span the possibilities still allowed
in the CMSSM, following the explorations made by LEP.  A grand summary of
the reaches of the various accelerators is presented graphically in
Fig.~\ref{fig:Manhattan}. The different levels of shading (color) present
the different types of sparticle: Higgses, charginos and neutralinos,
sleptons, squarks and gluino.  The first six points (I, L, B, G, C, J) are
presently favoured: they are compatible within 2 $\sigma$ with the present
$g_\mu - 2$ measurement, and the fine tuning is relatively small for most
of these points. Fig.~\ref{fig:Manhattan} summarizes the discussion of
this paper and exposes clearly the complementarity of hadron and electron
machines. It is apparent that many alternative scenarios need to be kept
in mind. 

Beyond the CMSSM framework we have discussed in this paper, one should
consider more general versions of the MSSM, in which the GUT universality 
assumption is relaxed~\cite{HinchBaer}. We note that
the phenomenology of {\it gauge-mediated} models~\cite{Dine} with
long-lived neutralino NLSP is very well covered by our suggested CMSSM
benchmark points, whilst a charged long-lived NLSP is found for our point
H.  Gauge-mediated models with a promptly-decaying NLSP give rise to
distinct signatures, but there are very few theoretical models that
predict these. The minimal {\it gaugino-mediated} models~\cite{Kaplan}
predict spectra~\cite{Schmaltz:2000ei} which are very similar to those of
the benchmark points in the `bulk' region discussed above, and are
therefore to a large extent covered by our analysis.  {\it Anomaly
mediation} provides an interesting framework~\cite{Randall} for model
building~\cite{Pomarol:1999ie}. A `phenomenological' model of anomaly
mediation~\cite{Gherghetta:1999sw} has been incorporated in {\tt
ISASUGRA}, and might serve as a basis for future studies. Finally, we
recall that {\it $R$-violating} models generally do not provide a suitable
dark matter candidate, and contain so many new $R$-violating couplings
that all the possibilities cannot be covered with only a few benchmark
choices. 

As we have discussed, the Tevatron has a chance to make the first inroads
into the spectroscopy of the CMSSM models we have studied, with the best
chance being offered by the lightest Higgs boson, followed by
chargino/neutralino searches.

The LHC is expected to observe at least one CMSSM Higgs boson in all
possible scenarios, and will in addition discover supersymmetry in most of
the models studied. However, we do observe that the discovery of
supersymmetry at the LHC is apparently not guaranteed, as exemplified by
benchmarks H and M. It would be valuable to explore the extent to which
precision measurements at the LHC could find indirect evidence for
supersymmetry in such scenarios. We note also that, in these cases, the
squarks and gluinos lie not far beyond the nominal physics reach of the
LHC, and an upgrade of the luminosity to 10$^{35}$ cm$^{-2}$s$^{-1}$ might
bring them within reach. 

An $e^+ e^-$ linear collider in the TeV range would in most cases bring
important additional discoveries, exceptions being benchmarks H and M, and
possibly E. Moreover, such a linear collider would also provide many
high-precision measurements of the Higgs boson and supersymmetric particle
masses and decay modes, that would play a pivotal r\^ole in first
checking the CMSSM assumptions and subsequently pinning down its
parameters, or those of the model that supplants it. As such, it will be
an essential tool for securing the supersymmetric revolution we
anticipate.

In many of the scenarios proposed, the discovery and exploration of the
complete set of supersymmetric particles, and especially some of the heavy
Higgses, gauginos and sleptons, will have to await the advent of a machine
like CLIC, which may need to run at an energy considerably higher than 3
TeV.  In particular, points F and M are very challenging and would need a
$\ell^+ \ell^-$ collider with a centre-of-mass energy of up to 7.5 TeV in
order to discover all the CMSSM particles. Distinguishing the different
squark flavours could be an interesting challenge for CLIC. 

If the CLIC technology cannot be extended this far, here might come a
r\^ole for a high-energy $\mu^+ \mu^-$ collider, if its neutrino radiation
problems could be overcome. A lower-energy $\mu^+ \mu^-$ collider would
also be interesting, with its unique capabilities for Higgs measurements. 

History reminds us that benchmarks have a limited shelf-life: at most one
of them can be correct, and most probably none. In the near future, the
CMSSM parameter space will be coming under increasing pressure from
improved measurements of $g_\mu - 2$, assuming that the present
theoretical understanding can also be improved, and $b \rightarrow s
\gamma$, where the $B$ factories will soon be dominating the measurements.
We also anticipate significant improvement in the sensitivity of
searches for supersymmetric dark matter.

We also note that astrophysical and cosmological estimates of the cold
dark matter density are converging, which may improve the upper limit on
$\Omega_\chi h^2$. As we mentioned earlier, lower values of $\ohsq$, even
below 0.1, would be possible if there exists some additional form of cold
dark matter. In this case, the sparticle spectrum might be somewhat
lighter than in the benchmarks we propose, though the scope for this is
rather limited. An interesting question for the future is whether the
sparticle measurements at the LHC and particularly a linear $e^+ e^-$
collider will enable accurate calculations of $\ohsq$ to be made
\cite{Drees:2001he,EO}.  This certainly seems possible in many of the
benchmarks we propose, though probably not for those with very large
values of $\Delta^\Omega$. 

Needless to say, the preliminary observations presented above need to be
confirmed by more detailed exploration of the benchmark scenarios we
propose, and of any others proposed within the context of different
supersymmetric model assumptions. These more detailed studies would
certainly benefit from improvements in the available simulation codes. All
of the necessary ingredients for a complete NLO analysis are already
available~\cite{Pierce:1997zz}. In order to achieve the necessary
precision, one now has to include the complete one-loop corrections to the
physical particle masses as well as perform a careful treatment of
sparticle thresholds, the effective potential and the Higgs boson mass
parameters.

\vskip 0.5in
\vbox{
\noindent{ {\bf Acknowledgments} } \\
\noindent  
The work of K.A.O. was supported partly by DOE grant
DE--FG02--94ER--40823. K.T.M. thanks the Fermilab Theory Group 
for hospitality during the completion of this work.}

\end{document}